\documentclass[a4paper,prb,twocolumn,showpacs,preprintnumbers,amsmath,amssymb,groupeaddress,superscriptaddress]{revtex4}
\usepackage{graphicx}
\usepackage{dcolumn}
\usepackage{bm}
\usepackage{nicefrac} 
\usepackage{amsmath}

\def\redtwo {}

\newcommand{\VEV}[1]{\left\langle {#1}\right\rangle} 
\newcommand{\abs}[1]{\left\vert {#1}\right\vert}

\def\kelvin{\hbox{\rm K}}

\def\mnt{Mn$^{3+}$}

\def\tesla{\hbox{\rm T}}

\def\LSMO{La$_{1-x}$Sr$_{x}$\-MnO$_3$}
\def\LCMO{La$_{1-x}$Ca$_{x}$\-MnO$_3$}

\def\LCSMO{La$_{1-x}($Ca$_y$Sr$_{1-y}$)$_{x}$\-MnO$_3$}
\def\oLCMO{La$_{5/8}$Ca$_{3/8}$\-MnO$_3$}
\def\oLSMO{La$_{5/8}$Sr$_{3/8}$\-MnO$_3$}
\def\oLCSMO{La$_{5/8}$(Ca$_y$Sr$_{1-y}$)$_{3/8}$\-MnO$_3$}
\def\uLSMO{La$_{0.75}$Sr$_{0.25}$\-MnO$_3$}
\def\uLCMO{La$_{0.7}$Ca$_{0.3}$\-MnO$_3$}
\def\uLCSMO{La$_{0.7}$(Ca$_y$Sr$_{1-y}$)$_{0.3}$\-MnO$_3$}

\newcommand{\commentout}[1]{}

\def\addrParma{Dipartimento di Fisica e Unit\`a CNISM, Universit\`a degli Studi di Parma, Viale G.\ Usberti 7A I-43100 Parma, Italy}
\def\addrPavia {Dipartimento di Fisica “A. Volta” e Unit\`a CNISM, 
Universit\`a degli Studi di Pavia, Via Bassi 6  I-27100 Pavia, Italy}

\def\addrDresden{Institute for Solid State Research, IFW Dresden, 
P.O.\ Box 27\,01\,16 D-01171 Dresden, Germany}

\def\addrWien{Institut f\"ur Festk\"orperphysik, Technische Universit\"at Wien, Wiedner Hauptstra\ss e 8-10 A-1040 Wien, Austria}
\def\addrCalestani{Dipartimento di Chimica Generale ed Inorganica, Universit\`a
di Parma, Viale G.\ Usberti 17A I-43100 Parma, Italy}

\begin{document}


\title{Band filling effect on polaron localization 
in La$_{1-x}($Ca$_y$Sr$_{1-y}$)$_{x}$MnO$_3$ manganites} 

\author{G. Allodi}
\email{Giuseppe.Allodi@fis.unipr.it}
\author{R. De Renzi}
\affiliation{\addrParma}
\author{\redtwo K. Zheng}
\affiliation{\addrParma}
\author{S. Sanna}
\affiliation{\addrPavia}
\author{A. Sidorenko}
\affiliation{\addrWien}
\author{C. Baumann}
\affiliation{\addrDresden} 
\author{L. Righi}
\author{F. Orlandi}
\author{G. Calestani}
\affiliation{\addrCalestani}

\date{\today}

\begin{abstract}

We report on an a $\mu$SR and $^{55}$Mn NMR investigation 
of the magnetic order 
parameter as a function of temperature 
in the 
optimally doped \oLCSMO\ and in the underdoped \LSMO\ and \LCMO\
metallic manganite families. The study is aimed at unraveling the 
effect of lattice distortions, implicitly  controlled by the  
Ca-Sr isoelectronic substitution, from that of hole doping $x$ on  the
Curie temperature $T_c$ and the order of the magnetic transition.

At optimal doping, the transitions are second order at all $y$ values, 
including
the $y=1$ (\oLCMO) end member. In contrast, they are first order in the 
underdoped samples, which show a finite (truncated) 
order parameter at the Curie point,
including \uLSMO\
whose $T_c$ is much higher than that of \oLCMO.
The order parameter 
curves, on the other hand, exhibit a very minor dependence on $x$, 
if truncation is excepted.
This suggests 
that the effective exchange interaction between Mn ions is  
essentially governed by local distortions, in agreement with the original 
double-exchange model, while truncation is primarily, if not entirely, an effect
of under- or overdoping.
A phase diagram, separating in the $x-y$ plane polaron-driven first order
transitions from regular second order transitions governed by critical 
fluctuations, is proposed for the \LCSMO\ system.
\end{abstract}


\maketitle
\section{Introduction}

Mixed valence manganites with the perovskite structure R$_{1-x}$A$_x$MnO$_3$ 
(R=lanthanide, A=alkali-earth metal) have been the subject of renewed interest 
since the discovery of a colossal magnetoresistance (CMR) at suitable 
compositions,
\cite{jin1994} probably the most spectacular manifestation of their 
half-metallic (i.e.\ spin-polarized) electronic transport.
The interplay between half metallic conduction and ferromagnetism is 
generally understood on the basis of the spin-conserving transfer
of an $e_g$ hole between two 
neighboring Mn ions, assisted by the ligand $p$ orbital,
referred to as double exchange (DE). 
\cite{zener} 
Due to the strong Hund-rule coupling between $e_g$ states and 
$t_{2g}$ core shells, the hole transfer is energetically favored only 
in the presence of a ferromagnetic (FM) order. In turn, DE 
can be viewed at as an effective ferromagnetic exchange interactions between 
manganese ions, mediated by mobile holes. Such an interaction competes with 
antiferromagnetic (AF) terms (i.e.\ superexchange) dominant in the undoped or 
underdoped materials, where they lead to AF insulating phases.  
Above a critical hole concentration $x$ (e.g.\ $x \ge 0.175$ in \LSMO), 
\cite{tokura_lsmo,phadia_lsmo_pcmo} 
double exchange dominates and drives the system 
into a FM-ordered half-metallic ground state. 
At finite temperatures, the fluctuating component of the magnetic moment 
behaves as a spin-flipping scattering term which reduces the polarization of 
charge carriers. The half-metallic band is disrupted at the Curie point $T_c$, which also coincides with a metal-insulator (MI) transition.

While the general agreement on the major role of double exchange in the 
physics of CMR manganites has never been challenged since 
Zener's work,\cite{zener}
the relevance of other interactions 
was stressed after their revival of interest  in the nineties. For instance, 
Millis {\it et al.}\cite{millis96} pointed out that a 
detailed description of the MI transition at $T_c$  requires taking into 
account the electron-lattice coupling of the \mnt\ ion, which is large in view 
of its Jahn-Teller character. Such a coupling term contributes to the 
effective mass of holes and favors their localization into Jahn-Teller 
polarons above $T_c$.

The critical temperature $T_c$ 
 varies with doping and exhibits, for each manganite family, a maximum at 
an optimal hole concentration $x \approx 3/8$. 
\cite{hwang_cheong, dagotto_revue} 
It strongly depends, however, also on the A-site cations. 
For instance, $T_c$ ranges from 
 370~K in optimally doped \LSMO\ (LSMO) down to approx.\ 268~K in \LCMO\ 
(LCMO), while the Pr-Ca series (PCMO) does not 
exhibit any metallic FM phase. \cite{phadia_lsmo_pcmo,phadia_lcmo}
The dependence of $T_c$ on the A-site isoelectronic substitution 
is understood in terms of 
the distortions and tilting of the MnO$_6$ octahedra (minimal in LSMO, 
maximal in PCMO) induced by 
the size mismatch of the cation.\cite{attfield}
Such distortions reduce the overlap of the Mn an ligand wave functions, 
either bending the Mn-O-M bond or 
increasing the Mn-O bonding distance. 
Within the framework of Zener's  model, this also
weakens the DE integral, 
resulting in a smaller 
electronic bandwidth and hence a lower $T_c$.

The nature itself of the magnetic transition apparently depends on the 
electronic bandwidth. Inelastic neutron scattering,\cite{lynn} 
$^{119}$Sn M\"ossbauer spectroscopy, \cite{mossbauer} 
and $^{139}$La nuclear magnetic resonance (NMR), \cite{gubkin} 
revealed a regular second order transition in nearly 
optimally doped LSMO and, 
on the other hand, "truncated" (i.e.\ first-order) 
ones in LCMO and other narrow-band manganites. 
The truncation effect is apparent from 
the dispersion curve of  magnetic neutron scattering, in the form of a spin 
wave stiffness 
which does not renormalize to zero at $T_c$.  \cite{lynn}
As seen from microscopic probes of magnetism 
like M\"ossbauer and NMR spectroscopies, which detect a hyperfine 
field directly proportional to the local magnetization, 
truncation shows up as 
a finite order parameter  and a phase-separated magnetic state at  the Curie 
point, whereby the volume of the ordered phase, rather than the order 
parameter, vanishes for $T \to T_c$. 
The first-order character of the transition indicates 
that the transition
 is not governed by critical
magnetic fluctuations. The large oxygen isotopic effect on $T_c$ in LCMO 
\cite{isotope_nature} demonstrates that lattice excitations, 
identified with Jahn-Teller polarons, 
 are involved. 
The spin wave dispersion, hence the exchange 
coupling, on the other hand, are however unaffected by the oxygen isotope 
substitution. \cite{lynn}
These observations suggest that 
the MI transition is driven by the formation of a polaron 
phase above $T_c$, which abruptly disrupts both 
the metallic state and the FM order.

The literature on the character of the 
magnetic transition, either first or second order, in CMR manganites, 
is nevertheless sparse.  
Moreover, each of the microscopic probes 
employed in such studies is affected by 
problems and limitations. For instance, 
 neutron scattering can access the nature of the transition only indirectly
through the spin wave stiffness, \cite{lynn} contrary to the 
local probes 
of magnetism, which detect directly a hyperfine field proportional to the order 
parameter. Among the latter, however,
M\"ossbauer spectroscopy requires substituting a 
non-magnetic ion at the Mn site, which perturbs 
appreciably the systems even at moderate substitutions, as it is apparent 
from the reduced $T_c$.\cite{mossbauer} 
On the other hand, zero field (ZF) NMR is not usually applicable in the 
critical 
region due to the drop of the resonance frequency and the divergence of the 
spin-spin relaxation rate $T_2^{-1}$ as $T_c$ is approached, unless the 
truncation effect on the transition is severe.\cite{gubkin}

In this paper we report an investigation of the order parameter in the 
optimally doped \oLCSMO\ (LCSMO) 
 isoelectronic series as a function of $y$. The Ca-Sr solid solution behaves
 as a 
virtual cation interpolating between Ca and Sr, and provides a 
continuous control over the induced lattice distortions, 
hence on the electronic bandwidth, at constant 
hole concentration. 
The two end members \LSMO\ and \LCMO\ are 
also investigated as a function of hole concentration $x$ in the underdoping 
regime, $0.2 < x < 3/8$. 
The comparison of the $x$ and $y$ dependences is aimed at unraveling the 
effect of charge on the order parameter and the type of magnetic transition, 
from those of distortions. For reference, the crystal cell parameters and 
Mn-O-Mn
bond angle are determined by X-ray diffraction.

The order parameter 
is studied as a function of 
temperature by combining ZF $^{55}$Mn NMR and muon spin rotation 
($\mu$SR).
The two techniques effectively complement
one another. 
$\mu$SR overcomes the aforementioned limits of NMR close to the 
magnetic transitions, but is afflicted in turn by line broadening 
and anomalies in the precession frequencies at lower temperature, seemingly 
related to frozen local disorder 
at the muon site. 
The isotropic and essentially on-site hyperfine field at $^{55}$Mn, on the 
contrary, is insensitive to such disorder and exhibits smooth temperature 
dependencies. The combination of $\mu$SR close to $T_c$ and NMR well below 
provides therefore a
 reliable microscopic determination of the order parameter.

We found second order transitions throughout the optimally doped 
LCSMO series and, in contrast, first order transitions in
underdoped metallic LCMO ($x \le 0.30$) and LSMO ($x \le 0.25$), despite the 
much higher $T_c$ in e.g.\ \uLSMO\ than in optimally 
doped \oLCMO. These findings, along with the scaling
properties of the order parameter curves, indicate 
that the effective exchange coupling is primarily governed by distortions
and the truncation phenomenon by charge 
doping.

The paper is organized as follows. 
Section \ref{sec:experiment} provides a brief description of sample 
preparation and  X-ray characterization, as well as 
of the experimental apparatus and methods.
The experimental results achieved 
by $\mu$SR, $^{55}$Mn NMR, and the 
combination of the two microscopic probes, are illustrated
in distinct subsections of Sec.\  \ref{sec:results}.
The effects of charge and lattice distortions the order parameter and the 
nature of the magnetic
transition, as they appear from the two local probes of magnetism, 
 are finally discussed in section \ref{sec:discussion}. 
Two appendices deal with the 
couplings of the two probes to the local magnetic moments 
in metallic manganites. 
The nature of the
 $x$-dependent hyperfine field at the $^{55}$Mn nucleus is dealt with 
in Appendix\ \ref{sec:coupling.nmr}, while the origin of the discrepancies
between the internal fields and the relaxations  experienced by muons and 
nuclei is discussed to some extent in Appendix \ref{sec:coupling.muon}.

\section{Experiment}
\label{sec:experiment}

\subsection{Sample preparation}
\label{sec:experiment.samples}

The polycrystalline samples were obtained by a standard solid-state
reaction from La$_2$O$_3$, CaCO$_3$, SrCO$_3$ and
  MnO$_2$. Reagents were dehydrated for several hours in a furnace at
150 C for one day. The La$_2$O$_3$ was further heated for several days at
1100 C in order to eliminate any trace of lanthanum hydroxide.
Reagents were ground by a Pulverisette~7 machine, weighed
in stoichiometric amounts in inert and dry atmosphere, pelletized,
fired in air at 1200 C for 12 hours in a furnace,  
 and finally cooled slowly to room temperature. The reaction products were 
then reground and subject  three times to the above thermal treatment - the 
last one 
at 1360 C for 20 hours - in order to ensure
a complete decomposition of the carbonates. 

The final characterization by X-ray powder diffraction 
revealed in the various samples a crystal structure belonging to the 
R$\bar 3$c or Pnma space groups, as 
appropriate of LSMO and LCMO, respectively. \oLCSMO\ with $0.5 <y<0.65$, 
showing 
a nanoscale admixture of the two structures, is a special case, discussed 
below. In any case, no  spurious phases was ever detected.

\subsection{X-ray characterization} 
\label{sec:experiment.xray}

Powder X-ray diffraction data from the LCSMO and LSMO samples
were collected at room temperature using 
Cu-Ka radiation with a Thermo Electron Corporation 
    X'tra\textsuperscript{\textregistered} diffractometer equipped with a Si(Li) energy selective detector. 
Diffraction 
patterns were acquired in the range $20^\circ \le 2\theta \le 120^\circ$
with a step of $0.02^\circ$. 
Structural refinement was carried out for each composition 
by the Rietveld method implemented in the GSAS package.\cite{gsas}  
The refinement 
was mostly aimed at determining 
the Mn-O bond lengths and Mn-O-Mn bond angles as a function of 
composition. In particular, we regard bond angles as the characteristic 
distortion parameters of the perovskite cell.

\begin{figure}
\includegraphics[width=\columnwidth]{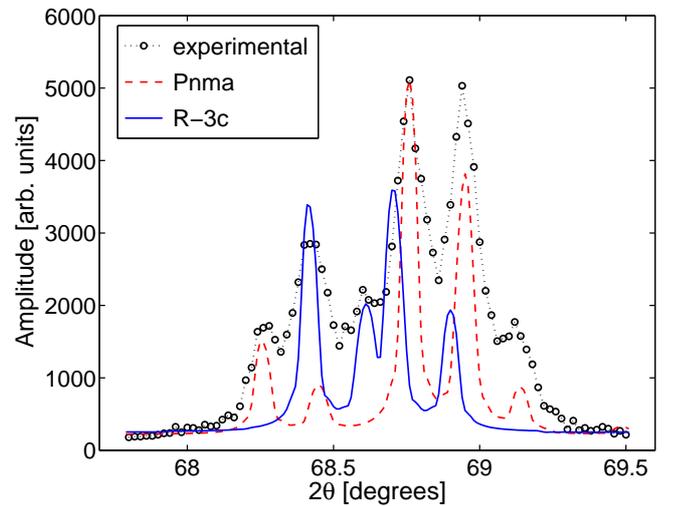}%
\caption{\label{fig:2theta}(color online) X-ray powder diffractogram 
at large angles from the $y=0.56$ LCSMO sample (symbols), and 
fitting patterns
according to the orthorhombic and rhombohedral systems (lines). }
\end{figure}

The Rietveld refinements yielded accurate fits of the X-ray powder diffraction 
patterns of \oLCSMO\ at the Sr-rich ($y \le 0.5$) and the Ca-rich ($y > 0.65$) 
ends of the 
phase diagram. Low-$y$ LCSMO exhibits the rhombohedral symmetry of the  
R$\bar 3$c space group, while the high-$y$ compositions belong to the 
orthorhombic system (Pnma space group), in agreement with the 
literature. 
At intermediate cation substitutions $0.5 < y < 0.65$, however, a single 
crystalline structure could not be refined. Typical diffraction data are 
shown in Fig.~\ref{fig:2theta} for a representative sample,
where a portion of a $2\theta$ scan is plotted, overlaid to the 
diffraction patterns fitted 
according to the R$\bar 3$c and Pnma space groups.   
The observed profile qualitatively coincides with the superposition of the 
R$\bar 3$c and Pnma patterns, which indicates the coexistence of the two 
structures. 
The poor quantitative agreement of a two-component fit, as well as 
the excess linewidth of the observed peaks,
however,
rule out a macroscopic phase mixture, and rather point to the spontaneous 
segregation of nanoscopic orthorhombic and rhombohedral clusters. 
The same nanoscale phase 
separation, 
with similar intermiscibility limits of Ca and Sr, was previously reported by 
Y. P. Lee {\it et al.\ } in \uLCSMO\ thin films. \cite{YP_Lee}

In the following, we restrict therefore our Rietveld 
analysis to the single-phase compositions 
of \oLCSMO\
and to underdoped LSMO, since the
near-overlap of reflections from the two coexisting structures
in LCSMO $0.5 < y < 0.65$, each one with similar 
unit cell parameters, prevented a reliable quantitative assessment. 
Soft restraints 
were applied to the mean Mn-O distance, in order to achieve fit convergence and to account for the weak 
contribution to X-ray 
scattering from oxygen atoms. Such restraints were tuned on the basis 
of the known structural 
properties of manganese perovskites reported in a vast literature.

\begin{figure}
\includegraphics[width=\columnwidth]{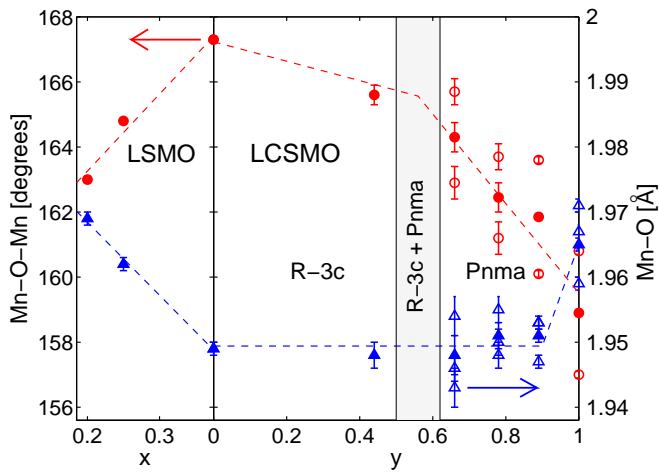}
\caption{\label{fig:rietveld}(color online) 
{\redtwo Best-fit Mn-O-Mn 
bond angles (bullets) and Mn-O bond lengths (triangles) 
in \LSMO\ as a function of $x$ (left panel) and 
in \oLCSMO\ as a function of $y$ (right panel). }
 The dashed lines are guides to the eye.
In the Pnma structures ($y>0.6$), angles and 
lengths of inequivalent bonds are marked by open symbols, while filled symbols 
mark average quantities. 
}
\end{figure}

The refined values for the 
Mn-O-Mn angles and Mn-O distances in LCSMO are plotted in  
Fig.~\ref{fig:rietveld}  as a function of 
$y$. In the orthorhombic phases, exhibiting inequivalent Mn-O bonds in the
unit cell, such quantities are split into multiplets (open symbols) by the  
the lower crystal symmetry. In this case, mean bond angles and lengths are also 
plotted (filled symbols), in order to allow a comparison of the cell 
parameters irrespective of the different crystal structures.
The figure clearly shows that in the LCSMO system the deviation from 
the Mn-O-Mn collinearity, i.e.\ the tilting of the 
MnO$_6$ octahedra, increases smoothly with increasing $y$, without any
apparent singularity at the transition from the R$\bar 3$c to the Pnma 
structure. Similarly, no clear anomaly is observed in the mean Mn-O 
bond length vs.\ $y$, if we except a steep upturn 
as the LCMO composition is approached.
The latter reflects the drop of $T_C$ below room temperature 
for $y\to 1$,  
hence the crossing of the FM-PM phase boundary in the present 
scan at room temperature,
which is also accompanied  by a marked step in the lattice 
parameters. \cite{Radaelli}

In the figure, 
the Mn-O-Mn and Mn-O bond angles and lengths in
the 
\LSMO\ samples as well are plotted vs.\ $x$ (left panel).   
A progressive decrease of the perovskite cell distortion and a 
contraction of the bond length are observed for $x$ increasing from 
underdoping  to optimal doping.
{\redtwo
This trend agrees with an increased A-site mean cation 
size, hence a reduced deviation of the 
Goldschmidt tolerance factor from unit, at higher Sr concentration. } 
\cite{cation_size}

\subsection{Magnetic characterization}
\label{sec:experiment.chirf}

The Curie temperature was measured in all samples by the simple 
radiofrequency (rf) apparatus 
described in Ref.\ \onlinecite{lpsmo2008}. A precise determination
of $T_c$ is also obtained directly from the longitudinal muon spin polarization 
data (see below), which are however available only  
for a subset of our samples. 
The rf method consists in recording the temperature-dependent inductance 
$L(T)$ of a small coil filled with the  
sample, powdered in order to minimize the diamagnetic shielding by 
eddy currents. 
The inductance depends on the volume magnetic rf susceptibility 
$\chi_{rf}$, the demagnetization coefficient $N$ ($N\approx 1/3$ for 
approximately spherical grains), and the filling factor $f$ 
of the coil, 
according to the relation
\begin{equation} 
L = L_0 [1 + f4\pi\chi_{rf}/(1+N4\pi\chi_{rf})]
\label{eq:chirf}
\end{equation} 
 where $L_0$ is inductance of the empty coil.  
$ \chi_{rf} \gg 1$ below $T_c$, given the soft ferromagnetism 
of the materials, and $L/L_0$ saturates to a value $ 1 + f/N$, in the order of 
units. We defined experimentally $T_c$ as the temperature 
at which $L(T)/L_0-1$ 
equals half the saturation value. Under the hypothesis of a spatial 
distribution of Curie temperatures, due to the unavoidable chemical disorder, 
and the approximation of a negligible contribution to $ \chi$ from the spatial 
components  whereby $T_c < T$,
such a definition 
corresponds to half the sample volume 
having undergone the magnetic transition. 
In the compounds subject to $\mu$SR experiments, the rf and 
$\mu$SR determinations of $T_c$ where 
found to coincide within the absolute accuracy of the thermometer calibrations,
 on the order of 1~K.

\subsection{Zero field $^{55}$Mn NMR}
\label{sec:experiment.nmr}
The hyperfine field $B_{hf}$ at the $^{55}$Mn nucleus, proportional to the 
thermal average of the
 Mn electronic moments, 
was measured vs.\ temperature by ZF $^{55}$Mn NMR. In the magnetically ordered 
phase, a spontaneous NMR signal is detected at a mean resonance 
frequency $^{55}\bar\nu = ^{55}\gamma/2\pi \times \bar B_{hf}$, 
where $^{55}\gamma/2\pi = 10.5$~MHz/T is the $^{55}$Mn gyromagnetic ratio.
The 
experiments were performed in zero field by means of a 
home-built phase-coherent spectrometer \cite{hyrespect} and a helium-flow 
(in the 2-70~K range) or a nitrogen-flow cryostat (above 70~K). Specimens of 
typical 
mass 20~mg, finely ground in order to maximize the penetration of the rf field,
were placed in 
an untuned probehead consisting of a small coil 
($\approx 30$~nH) terminated onto a 50~$\Omega$ resistor. 
The reduced sensitivity 
and rf field provided by the resistive circuit were overcompensated by 
the large 
rf enhancement, of the order of thousands, characteristic of the NMR signals 
from domain walls in ferromagnets. \cite{riedi_eta, turov}
The usage of a non-resonant circuit, on the other hand, allowed the 
automation of unattended frequency scans.
A conventional tuned LC circuit was sometimes employed only 
very close to $T_c$, in order to improve sensitivity 
and approach the transition as close as possible.

The inhomogeneously broadened spectra were recorded point by point 
by exciting spin echoes at discrete frequencies by means of a standard 
$P-\tau-P$ pulse sequence, with delays $\tau$ of 2.5-4 
$\mu s$, and equal rf pulses $P$ of 0.3-1.4 $\mu s$ with intensity optimized 
for maximum signal. The whole sequence was kept as short as possible compatibly
with 
the dead time of apparatus, especially at high temperature, 
in view of the very short $T_2$ relaxation times,  of the order of 1~$\mu s$
 close to $T_c$. 
The digitized echo signals were analyzed by the fast Fourier transform (FFT), 
assigning the maximum magnitude of the transformed 
echo to the spectral amplitude at the working frequency.  
This criterion 
 is preferred in practice over the textbook method of assigning the 
zero-shift FFT 
component, \cite{clark} 
as it partly compensates for the possible spectral 
hole burning at the center of the irradiated band. 
Such an artifact, due to
overdriven spin echoes, is often encountered in the NMR of ferromagnets,
where it usually originates from the spatial and/or spectral inhomogeneity of 
the enhancement factor.

\subsection{$\mu$SR experiments}
\label{sec:experiment.musr}

Zero-field $\mu$SR experiments 
on the LCSMO series 
were performed at the Laboratory for Muon Spin Spectroscopy (LMU) 
at Paul Scherrer Institut
(Villigen, CH) on the GPS spectrometer equipped with a closed cycle 
refrigerator (5-400~K) as a sample environment.  
ZF $\mu$SR data from the underdoped LCMO samples 
were also recorded on the MUSR instrument at ISIS (Chilton, UK). 
The observed quantity in a ZF $\mu$SR experiment is the time evolution of the 
asymmetry of the muon decay, $A(t)$, recorded in a pair of opposite positron 
detectors aligned parallel to the initial muon spin direction $\hat z$.    
Such a quantity is directly proportional to the projection  of the 
muon spin polarization $\bm{P}(t)$ along $\hat z$, while the orthogonal 
projections vanish identically 
due to the axial symmetry of the setup. \cite{Schenck}

In the paramagnetic (PM) phase, implanted muons experience no net 
magnetic field, 
and the asymmetry signal exhibits a pure longitudinal decay with a 
moderate decay rate $\lambda$, 
$A(t) =A_0 \exp(-\lambda t)$, 
where $A_0$ is the maximum asymmetry characteristic of the instrument setup.
In the ordered phase of a magnetic material, on the other hand, a spontaneous 
field sets in at the muon site, originating from the dipolar and hyperfine 
coupling of $\mu^+$ with the 
electronic moments. 
The muon may stop in principle at several magnetically inequivalent sites, 
each of them experiencing distinct internal fields $\bm{B}_{\mu}$, hence giving 
rise to distinct muon spin precessions.
For an arbitrary orientation of $\bm{B}_{\mu}$ with 
respect to the initial spin direction $\hat z$, however, also a non-precessing 
component (referred to as longitudinal), proportional 
in amplitude to the projection of  $\bm{B}_{\mu}$ along $\hat z$, 
shows up in the asymmetry signal.
The muon polarization $P_z(t) \equiv A(t)/A_0$ in the 
ordered phase is therefore fitted  as
 
\begin{equation}
P_z(t) =  \sum_{i=1}^m a_{Li} e^{-t/T_{1i}} +  
\sum_{j=1}^na_{Tj} e^{-\sigma_j^2t^2/2} \cos \omega_j t
\label{eq:asym}
\end{equation}
where $a_{Li}$, $T_{1i}$ are the 
amplitude and relaxation time of the longitudinal components,
$\omega_j\equiv\gamma_\mu B_{\mu\,j}$, $a_{Tj}$, 
$\sigma_j$ are the frequency, 
amplitude, and 
inhomogeneous Gaussian linewidth of the muon 
precession at the $j$-th site, respectively, and
$\gamma_\mu/2\pi$ = 135.54 MHz/T is the muon gyromagnetic ratio.
Here we account for spatially inhomogeneous longitudinal relaxations by 
including $n$ distinct decay terms (not necessarily related
 to any of the $m$ sites in the magnetic structure).

In a polycrystalline sample, cumulative weights $\sum_{i} a_{Li}=1/3$, 
$\sum_{j} a_{Tj}=2/3$ are predicted from the angular average of the internal 
fields.
The transverse signals, however, may be partly or totally lost in the case of 
very broad lines, hence experimentally $0 \le \sum_{j} a_{Tj}\le 2/3$.
In particular, they are completely lost at ISIS, due to the small bandwidth of 
the MUSR spectrometer operating with a pulsed muon beam. 
Similarly, $\sum_{i} a_{Li} < 1/3$ if a longitudinal component decays with 
an exceedingly large relaxation rate $\lambda = T_1^{-1}$.  
The total longitudinal amplitude $\sum_{i} a_{Li}$, on the other hand, may 
exceed $1/3$ in a phase-separated state with a magnetically ordered and a PM volume fractions, e.g.\ close to the transition in the presence of a 
distribution of $T_c$.




\section{Experimental results}
\label{sec:results} 
We report separately the results 
obtained in the various compounds by the microscopic probes of magnetism: $\mu$SR, NMR, and the combination 
of the two  techniques.

\subsection{$\mu$SR results}
\label{sec:results.musr}

We describe in some detail 
the $\mu$SR experiments carried out at LMU-PSI on optimally doped \oLCSMO,
accessing muon spin precessions.  
The  $\mu$SR spectra are qualitatively very similar in all the 
investigated samples ($y=0$, 0.33, 0.44, 0.5, 0.56, 0.67, 1). 
Their common 
features are summarized as follows.

\subsubsection{Muon spin precessions}
\begin{figure}
\includegraphics[width=\columnwidth]{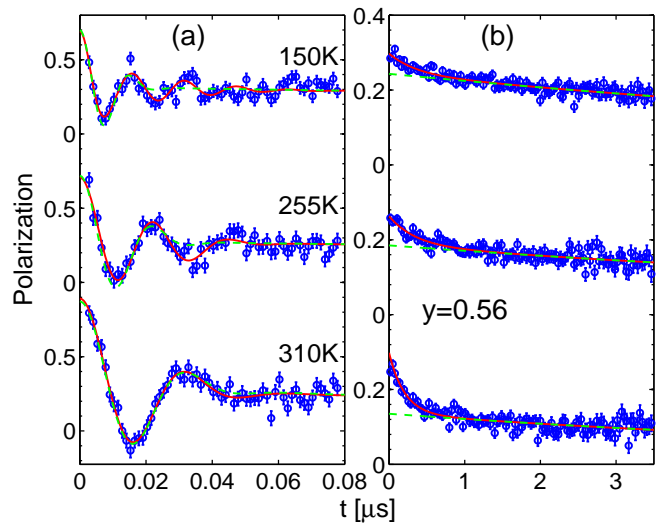}
\caption{\label{fig:time_diff}(color online) Time evolution of the 
muon spin polarization in LCSMO ($y=0.56$, $T_c=335$~K) at 310~K, 255~K, and 150~K. 
(a) Polarization vs.\ $t$ at early times. Dashed and  solid lines are fits to  
one and two precession components, respectively, plus a 
longitudinal decay.  
b) Longitudinal signals, 
with  the fast
precession components  
filtered out by binning counts at 30~ns resolution.
Solid lines are fits to a two-exponential decay; dashed lines mark the 
slower decay components.
}
\end{figure}

The time-differential muon polarization $P(t)$ is plotted at early times in 
Fig.~\ref{fig:time_diff}a for a representative sample ($y=0.56$) 
at selected 
temperatures.

The transverse signal in the FM phase is best fitted to a doublet of
 nearly degenerate 
precession components, 
with a frequency difference 
below the resolution limit (i.e.\ $\abs{\omega_1-\omega_2} < \sigma_j^2$).
 The $\chi^2$ improvement obtained with a two-component fit is however marginal
in an interval of several tens kelvin just below $T_c$, where a fit to a 
single damped precession is 
nearly as accurate.
In the same temperature range, 
the transverse polarization, extrapolated to the initial time $t=0$, 
equals within errors the theoretical
2/3  value expected for a polycrystalline sample, 
i.e.\ 
all muon sites and all the sample 
volume contribute to the transverse signal.

At lower temperatures, however, deviations from a 
single damped cosine function 
are appreciable. On further decreasing temperature, a reduction in the total  
transverse asymmetry is also observed. 
The missing fraction corresponds to an overdamped precession component, 
which is not detected due to a decay time shorter than the dead time of the 
GPS spectrometer ($\approx 5~$ns). 
Such a behavior,  
namely, the appearance of complex precession signals and a missing transverse fraction well below $T_c$, is somewhat
more marked on the low-$y$ side of the LCSMO series. In 
LSMO ($T_c \approx 300~\kelvin$), in particular, 
precession features are only visible above 300~K, while the signals 
observed at $T < 300~\kelvin$ are pure longitudinal decays. 

\begin{figure}
\includegraphics[width=\columnwidth]{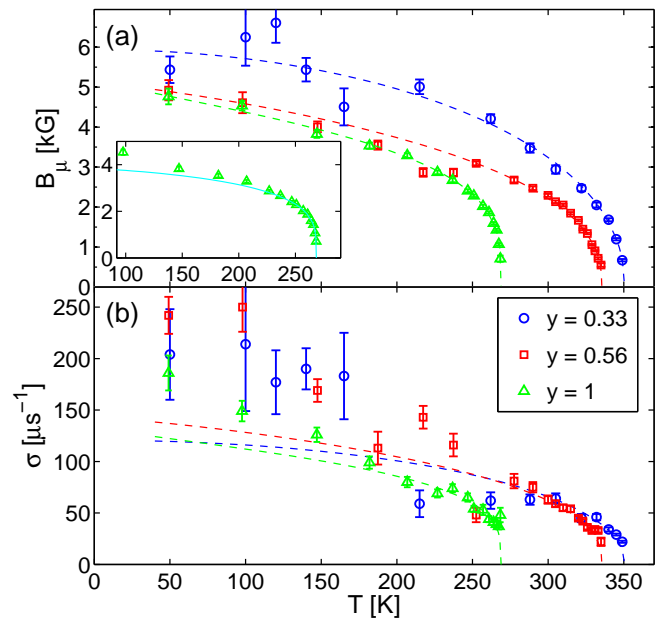}
\caption{\label{fig:mu_freq_sigma}(color online) Muon precession frequencies, 
in field units (a) and Gaussian  linewidths (b) vs.\ temperature of  three 
LCSMO samples: $y=0.33$ (bullets),$y=0.56$ (squares), and $y=1$ (triangles).
The dashed lines in panel (a) are guides to the eye. The same lines are 
reproduced in panel (b) rescaled by constant factors. 
Inset of panel (a): constrained 
fit of $B_\mu(T)$ ($y=1$) to Eq.\ \protect{\ref{eq:mlaw}} (see text).   
}
\end{figure}

The main quantities extracted from the muon precession spectra, namely, the 
internal field $B_\mu$ and the Gaussian precession linewidth $\sigma_\mu$,
are plotted 
vs.\ temperature in Fig.~\ref{fig:mu_freq_sigma} for representative 
compounds.
The data are obtained from a fit to a single transverse component as  
in Ref.\ \onlinecite{heffner} on LCMO, in spite of 
its non-perfect accuracy, as it yields smoother temperature dependences. 
In a two-component fit, the fields at the two sites $B_{\mu1}$, $B_{\mu2}$ are in fact highly correlated, and the resulting values are strongly scattered. 
Similarly, the precession linewidths and 
amplitudes are too highly correlated to be fitted independently  
in the present case of overdamped signals. We therefore 
fixed amplitudes and took the linewidth as a free-running parameter.

Low-temperature anomalies 
are observed in $B_\mu(T)$. 
A constrained fit  
to the empirical law 
\begin{equation} 
\nu(T)=\nu_0\left [1- \left (T/T_c^\star \right )^\alpha\right ]^\beta
\label{eq:mlaw}
\end{equation} 
interpolating between a low-temperature behavior 
$\nu_0 - \nu\propto T^\alpha$ and the critical power-law dependence 
$\nu\propto (1 -T/T_c^\star)^\beta$ of the order parameter,\cite{blundell}
 with the 
 $\alpha$ and $\beta$ exponents forced to take physically meaningful values
(see Sec.\ \ref{sec:results.nmr} and \ref{sec:results.ordparam}),  
reveals in fact an 
  upturn in $B_\mu(T)$ at approximately $T= 3 T_c/4$ (inset of  
Fig.~\ref{fig:mu_freq_sigma}a for $y=1$). 
Below the same temperatures, the 
departure of the precession patterns from a single damped cosine wave are 
more marked, and a two-component fit becomes significantly more accurate. 
Such a change in the precession waveforms is also signalled by a steep 
increase of the effective linewidths $\sigma_\mu(T)$ on cooling.
The dashed  lines overlaid to the data in the main panel of 
Fig.~\ref{fig:mu_freq_sigma}a 
are guides to the eye devoid of any physical significance, which are 
replicated, multiplied by constant factors, 
in Fig.~\ref{fig:mu_freq_sigma}b to 
facilitate the comparison of $B_\mu$ and $\sigma_\mu$. It is apparent that the 
two quantities do not scale with each other {\redtwo over the full 
temperature range}.
The relative 
 linewidths $\gamma_\mu^{-1}\sigma_\mu(T)/B_\mu(T)$ tend in fact to 
asymptotic values of approximately 1/4 for $T\to T_c$ and 
of 1/2 or higher for $T\to 0$.
Such an excess linewidth is 
 indicative of 
extra static disorder at low temperatures.

\subsubsection{Critical region}

\begin{figure}
\includegraphics[width=\columnwidth]{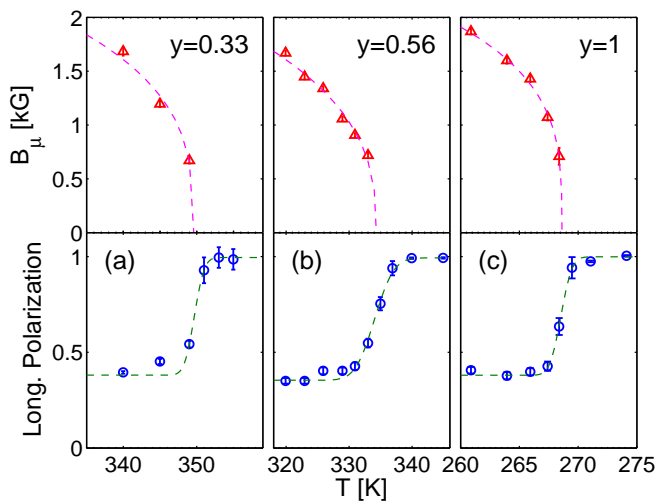}
\caption{\label{fig:mu_critical}(color online) Top: muon internal field for 
$T\to T_c$ in the $y=0.33$, $y=0.33$, and $y=1$ 
LCSMO samples (from left to right, respectively). The dashed lines are guides to the eye.
Bottom: total longitudinal polarization across the Curie transition in the 
three samples. The dashed lines are fits to Gaussian distributions  of 
$T_c$.}
\end{figure}

For clarity, we replot $B_\mu$ on an expanded temperature scale  
across the critical region in 
Fig.~\ref{fig:mu_critical}, overlaid to the total muon longitudinal amplitude 
$P_l=\sum_{i} a_{Li}$ (see Eq.~\ref{eq:asym}).
The transition from the FM phase to a fully PM state is marked by the 
recovery of the full longitudinal polarization.
The rise of $P_l$ to unit takes place over finite 
temperature intervals, indicating spatially inhomogeneous Curie 
temperatures in the samples. Fits of $P_l(T)$ to a Gaussian distribution of 
$T_c$ yield 
half-widths $\Delta T_c\approx 1$~K in the end members
LSMO and LCMO, up to $\Delta T_c\approx 3$~K in \oLCSMO\ with $1/3 < y < 2/3$. 
The larger $\Delta T_c$ in the Ca-Sr alloys
agree with their expected larger chemical inhomogeneities, 
due to local deviations of both $y$ and the hole doping $x= 3/8$ 
from the nominal mean values. Relative $T_c$ indeterminacies on this order of 
magnitude are however 
standard in non-stoichiometric magnetic oxides. 
For instance, even larger $\Delta T_c /T_c$  has been reported for
state-of-the-art samples of  other doped manganites. \cite{lpsmo2008}. 

Well-defined precessions 
(i.e., $\omega_\mu > \sigma_\mu$) are detected for all $y$ values 
up to temperatures $T_{max}$ corresponding to
reduced frequencies $\omega_\mu(T_{max})/\omega_\mu(0)\approx 0.1$. 
The precession frequencies 
 extrapolate to zero at
$T_c^\star\approx T_{max} + 1\,\kelvin$, a temperature reasonably well 
determined independent of 
the exact functional form of $\omega_\mu(T)$. It is apparent from the figure 
that 
$T_c^\star$ lies in the middle of 
the rising edge of $P_l(T)$ identifying the mean $T_c$, 
according to the above discussion. 
We conclude therefore that
 $T_c^\star = T_c$ all over the \oLCSMO\ 
series, including the LCMO end member, as in a regular second transition.

\subsubsection{Longitudinal signals}

We describe here the peculiar non-exponential longitudinal muon spin 
relaxations for sake of completeness, although of marginal interest to 
the main subject of this article. 
{\redtwo 
These results are discussed briefly at the end of the paper
in the context of  
 the muon coupling. }
 
The longitudinal polarization of a representative sample ($y=0.56$) 
is plotted vs.\ time 
in Fig.~\ref{fig:time_diff}b at a few temperatures below $T_c$,
with the precession components filtered out for clarity.
The presence of two decay components at $T\ge 100$~K, with decay constants 
$\lambda$ of the order of tens and tenths of reciprocal microseconds, 
respectively, is apparent in the figure. 
The same two-component behavior was reported in slightly underdoped LCMO by  
Heffner {\it et al.}. \cite{heffner2000}
Relaxations at $T\le 50$~K, on the contrary, are single exponential with a very small rate.

\begin{figure}
\includegraphics[width=\columnwidth]{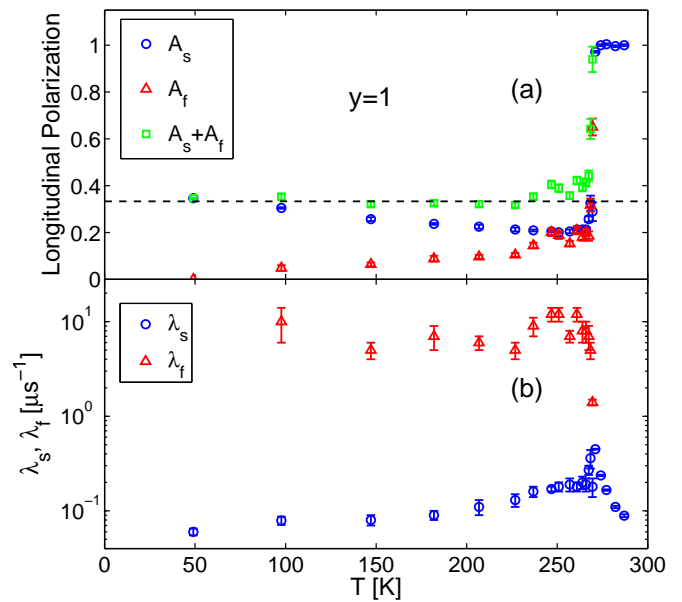}
\caption{\label{fig:mu_T1}(color online) Initial muon spin polarization $A_S$, $A_f$ 
(a) and relaxation rates 
$\lambda_s$, $\lambda_f$ (b)
of the slow- (bullets) and fast-relaxing (triangles) longitudinal 
fractions  in \oLCMO. 
The dashed horizontal line in panel (a) marks the $\nicefrac{1}{3}$ value appropriate 
for the total longitudinal polarization $A_S+A_f$ in a
polycrystalline sample below $T_c$.
}
\end{figure}

The decay rates and relative weights of the two decay components 
{\redtwo  of another sample ($y=1$)} are plotted 
vs.\ temperature in Fig.~\ref{fig:mu_T1}. Their temperature dependence is 
qualitatively identical 
across the LCSMO series and in underdoped \LCMO\ ($x$ = 0.25, 0.3). 
The amplitude $A_f$ of the 
fast-relaxing component progressively grows on warming above 100~K, up to a 
limit value of approximately 1/6 (i.e.\ half the magnitude of the 
longitudinal polarization  
in the ordered phase) for $T\to T_c$, while the amplitude $A_s$ of the 
slow-relaxing component follows a complementary behavior $ A_s\approx 1/3-A_f$.
The corresponding relaxation rates $\lambda_s$, $\lambda_f$, on the other 
hand, are nearly 
independent of $T$ over the same temperature interval, if we except for a 
shallow peak in $\lambda_s$ and  a 
slight decrease, instead of a critical divergence, in $\lambda_f$ for 
$T\to T_c$.

\subsection{Zero-field $^{55}$Mn NMR}
\label{sec:results.nmr}

\begin{figure}
\includegraphics[width=\columnwidth]{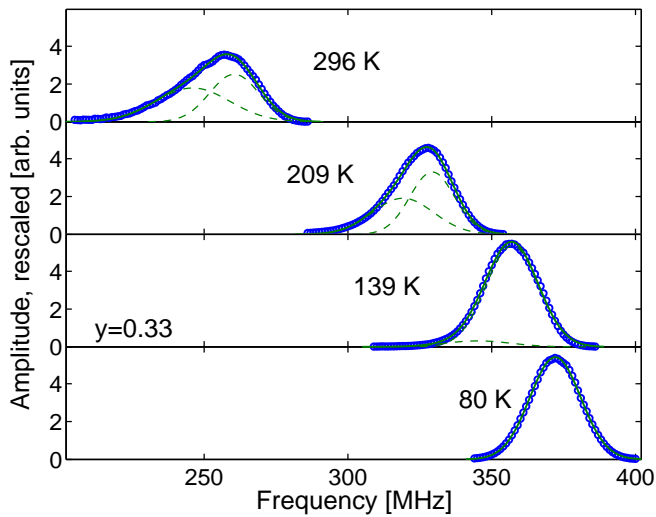}
\caption{\label{fig:nmrspectra}(color online) ZF $^{55}$Mn NMR spectra 
of LCSMO $y=0.33$ at different temperatures. The spectra are corrected 
in amplitude for 
the frequency dependent sensitivity $\propto \nu^2$ and scaled for clarity
by arbitrary vertical factors relative to each other. 
Solid and dashed lines are one- 
or two-Gaussian fits and individual Gaussian components, respectively. }
\end{figure}

All the compounds, the optimally doped Ca-Sr alloys as well as well as the 
LSMO and LCMO series, were subject to our $^{55}$Mn NMR study.
Very intense 
$^{55}$Mn spin echoes were excited in zero external field by applying very low 
rf power thanks to 
a sizable enhancement $\eta \approx 1000$, 
\cite{riedi_eta} a value assigning the 
signals to nuclei in domain walls. 
$^{55}$Mn spectra at several temperatures are plotted in 
Fig.~\ref{fig:nmrspectra} for a representative sample. 
 They are qualitatively very similar in all compounds.

Well below $T_c$, the $^{55}$Mn spectra are motionally narrowed 
nearly-Gaussian single peaks characteristic of a DE state of the Mn ion, 
\cite{matsumoto, savosta1998, our_films}
with low-temperature mean resonance 
frequencies $\bar\nu_{55}\approx 370$-390~MHz and half widths  
$\sigma\approx 8$-9~MHz.  
At $T > 100$~K, however, the spectra depart from the Gaussian shape, and at 
higher temperature, they clearly develop a low-frequency shoulder. 
Perfectly identical features were reported in metallic LCMO and LSMO by 
Savosta {\it et al.},\cite{savosta} 
who interpreted this behavior in terms of a subtle 
phase segregation on a nanoscopic scale, with slightly different electronic 
properties in 
the volume fraction probed by the spectral shoulder. 
Here, we are not going to rely on such a phase separation model, 
which is outside the scope of this paper. 
A reliable spectral decomposition into two or more 
unresolved peaks, on the other hand, depends critically on 
on the flatness in the frequency response of the receiver, as well as on an 
accurate calibration of the optimum excitation power vs.\ frequency, which are 
both difficult to attain over very broad spectra.
Therefore, we simply calculated a mean resonance frequency $^{55}\bar\nu$ 
as the weighted average of the peak centers in a two-component fit. 
The so-determined $^{55}\bar\nu$ follow smoother temperature dependences than 
the frequencies ${^{55}}\bar\nu_h > {{^{55}}\bar\nu}_l$ of the individual peaks. 
The indetermination of the order parameter estimate, related to the 
definition of a mean effective NMR frequency, 
is anyway 
unimportant and it does not affect significantly the following analysis.
For instance, $^{55}\bar\nu$ differs from the the position 
${^{55}}\bar\nu_h$ of the 
sharper peak (which could be taken as an alternative definition of a 
mean resonance frequency) by no more the 4\% in relative terms, i.e.\ 
 less than $\Delta \nu_\mu / \nu_\mu$,
except very close to $T_c$. 
Over the critical temperature interval, on the other hand,  our study of the 
order parameter  in LCSMO
mostly relies on the $\mu$SR frequency $\nu_\mu$.

\begin{figure}
\includegraphics[width=\columnwidth]{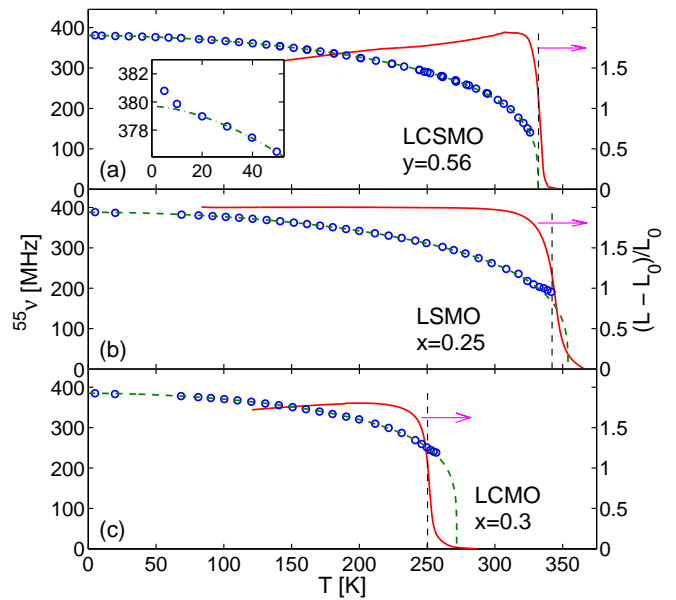}
\caption{\label{fig:NMR_vs_T}(color online) Mean NMR frequencies vs.\ $T$ 
(symbols), overlaid to fits to Eq.\ \protect{\ref{eq:mlaw}} (dashed lines), 
and normalized rf inductance $L(T)/L_0 -1$ (solid lines)  
in La$_{5/8}$(Ca$_{0.56}$Sr$_{0.44}$)$_{3/8}$\-MnO$_3$ (a),
La$_{0.75}$Sr$_{0.25}$\-MnO$_3$ (b), and La$_{0.7}$Ca$_{0.3}$\-MnO$_3$ (c).
The vertical dashed lines mark $T_c$, determined as the center of the 
edge of $L(T)$. Inset of panel (a): detail of $^{55}\nu(T)$ at low temperature 
(the dash-dotted line is a guide to the eye). }
\end{figure}

Typical $^{55}\nu(T)$ curves (henceforth we drop the bar over $\nu$) are 
plotted in Fig.~\ref{fig:NMR_vs_T} for a 
representative optimally doped \oLCSMO\ compound, and in 
Fig.~\ref{fig:NMR_vs_T}b,c for two
underdoped \LCMO\ and \LSMO\ samples, respectively.
In the figures, solid lines are fits to the phenomenological function 
of Eq.~\ref{eq:mlaw}.
Unlike the $\mu$SR data, the experimental $^{55}\nu(T)$ points follow regular 
curves without any anomalies
at intermediate temperatures, as it is witnessed by  the good 
accuracy 
of the 
fits to Eq.\ \ref{eq:mlaw}. The best fit parameters 
are tabulated in Tab.~\ref{tab:mlaw_pars} for the different compounds.
Overlaid to $^{55}\nu$, the figure plots for comparison also the rf magnetic 
response
$L(T)/L_0 - 1$ (see Sec.\ \ref{sec:experiment.chirf}), whose rising edge marks 
$T_c$. 

The truncated character of the FM transition is apparent 
in underdoped \LSMO\ ($x\le 0.25$) and 
\LCMO\ ($x\le 0.3$), which show finite resonance frequencies at $T_c$ and  
$^{55}\nu(T)$ curves extrapolating to zero at temperatures $T_c^\star$ larger 
than $T_c$ by 11~K (LSMO) or more. 
These samples  also exhibit $^{55}$Mn signals 
 up to several kelvin above the nominal $T_c$,
with corresponding mean frequencies $^{55}\nu$ clearly exceeding the values 
extrapolated from Eq.\ \ref{eq:mlaw}. Such a 
behavior indicates a phase separated state over a wide temperature interval 
across the transition, the wider the farther from optimal doping,
hence governed by chemical inhomogeneities. 

In the optimally doped LCSMO sample series, on the other hand, spectra can be 
recorded up to approximately $T_c-10$~K, corresponding to reduced  
temperatures $(1 -T/T_c)\approx 0.03$ and reduced order parameters 
$\nu(T)/\nu(0)\approx 0.4$. 
 Above such limiting temperatures, the spin-spin 
relaxation time $T_2$ falls below 1~$\mu$s, much shorter than the dead time of 
the NMR apparatus (2.5~$\mu$s), making spin echoes no longer detectable. 
The extrapolated $^{55}\nu(T)$ curve vanish at temperatures 
$T_c^\star$ compatible with $T_c$, 
in agreement with the behavior of $\nu_\mu(T)$.
We defer the analysis of the critical behavior to the next session, combining 
$\mu$SR and NMR data.

The inset of Fig.~\ref{fig:NMR_vs_T}a 
shows a  blow-up 
of the $^{55}\nu(T)$ plot  at low temperature.  The points at $T \le  
10$~K exhibit a clear upturn
relative to the best fits to Eq.\ \ref{eq:mlaw} by approximately 0.5-1~MHz. 
Similar low-temperature anomalies are systematically observed in the 
spectra of all our bulk materials, 
irrespective of the dopant species and doping concentration. They 
are absent, in contrast, in thin films of CMR manganites. \cite{unpublished}   
We have no clear explanation for this behavior, seemingly related to a {\redtwo
subtle freezing phenomenon. \cite{note_frequencypulling}}

\begin{figure}
\includegraphics[width=\columnwidth]{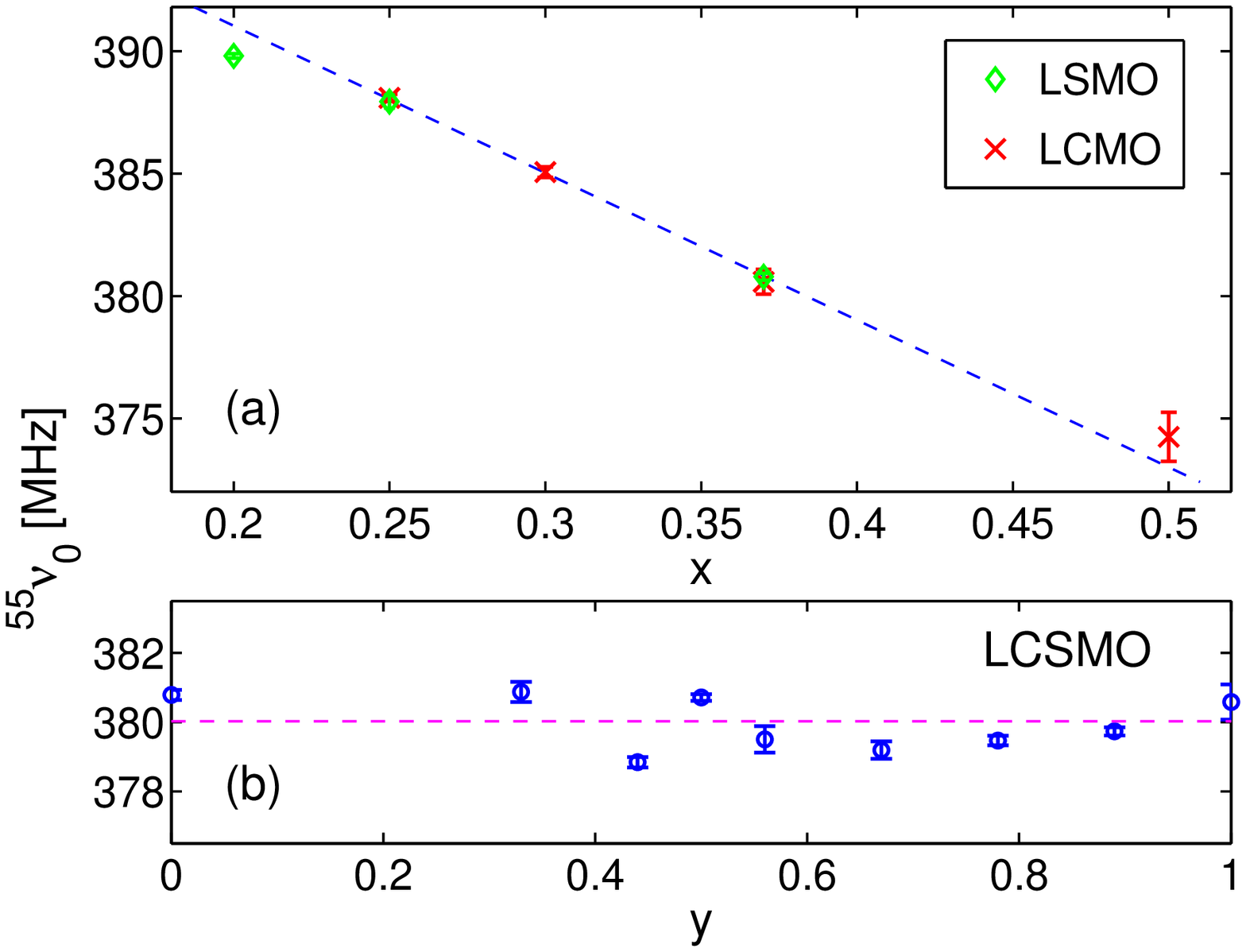}
\caption{\label{fig:NMR_nu0}(color online) Zero-temperature NMR frequency 
vs.\ $x$ (top) in \LSMO\ (diamonds) 
and \LCMO\ (crosses), and 
vs.\ $y$  in \oLCSMO\ (bottom). The LCMO $x=0.5$ point is taken from 
Ref.\ \protect{\onlinecite{allodi_ca50}}. 
The dashed lines 
in panels (a) and (b)
are fits to a straight line dependence and a constant, respectively.}
\end{figure}

The mean resonance frequencies $^{55}\nu_0$ at $T=0$, 
extrapolated from the fits to Eq.\ \ref{eq:mlaw}, are plotted 
vs.\  $x$ in Fig.~\ref{fig:NMR_nu0}a for the \LCMO, \LSMO\ families, and vs.\  $y$ for the \oLCSMO\ series in Fig.~\ref{fig:NMR_nu0}b. 
The datum of the overdoped LCMO sample ($x=0.5$) is taken from 
Ref.\ \onlinecite{allodi_ca50}.
$^{55}\nu_0$ follows a straight line dependence 
\begin{equation} 
^{55}\nu_0(x)=ax+b
\label{eq:straight}
\end{equation} 

with reasonable accuracy (Fig.~\ref{fig:NMR_nu0}a), 
with the same best-fit coefficients $a=-60\pm 2$ MHz, $b=403\pm 2$ MHz for the 
two dopant ions.  The independence of $^{55}\nu_0$ on the doping mechanism 
is consistent with Fig. \ref{fig:NMR_nu0}b, showing 
$^{55}\nu_0(y)=380\pm 1$~MHz approximately constant over the optimally doped 
Ca-Sr series. Assuming the independence on $y$ as exact, 
the 
deviation $\Delta^{55}\nu_0$ then provides an estimate for the deviation of $x$ 
from its nominal $x=3/8$ value.
Based on the empirical calibration given by the $a$ coefficient, we get  
$\Delta x\approx 0.015$, in agreement with the known accuracy of the 
preparation process.

\subsection{Order parameter}
\label{sec:results.ordparam}

We combine here the $\mu$SR and NMR techniques in an analysis of the 
temperature dependence of the order parameter in the optimally 
doped \oLCSMO\ series, with the muon and $^{55}$Mn 
precession frequencies constrained to the same temperature dependence.  
The $\mu$SR data are restricted to temperatures down to a few tens 
kelvin below $T_c$, a range overlapping the 
one accessed by NMR  over an interval of 20-30~K, where the two quantities 
are in an approximately constant ratio.
At lower temperature, the normalized $\nu_\mu(T)$ departs from $^{55}\nu(T)$, 
as pointed out above. 
In this case, however, $^{55}$Mn nuclei probe the local 
electronic moment more reliably, 
due to the essentially isotropic and on-site nature of their hyperfine 
coupling (see  Appendix \ref{sec:coupling.nmr}). 
Therefore, the $\mu$SR points at lower temperature 
are rejected from the fit. The 
reason for the discrepancy between the two probes is discussed  
in Appendix \ref{sec:coupling.muon}. 
The merger of $^{55}\nu(T)$ with the  
$\nu_\mu(T)$ curve, scaled in frequency by an empirical factor and limited 
to high temperatures as specified above,
represents our best determination of the order parameter by local probes 
of magnetism.

\begin{table*}
\caption{\label{tab:mlaw_pars} 
Fitting parameters to 
Eq.\ \protect{\ref{eq:mlaw}} of the 
$^{55}\nu(T)$  and combined $^{55}\nu(T)$\,-\,$\nu_\mu(T)$ data, and Curie temperatures determined 
from longitudinal $\mu$SR fractions and rf magnetic response, in the various 
samples subject of this work. 
}
\begin{ruledtabular}
\begin{tabular}{llllllllll} 
Sample  & \mbox{$^{55}\nu_0$ (MHz)} 
& \mbox{$B_{\mu}^{\circ}$}\,(kG)\footnotemark[1] & 
\mbox{$T_c^\star$ (K)\footnotemark[2]} &\mbox{$T_c^\star$ (K)\footnotemark[3]} 
& $\alpha$
& \mbox{$\beta$\footnotemark[2]} & \mbox{$\beta$\footnotemark[3]} 
& \mbox{$T_c$ (K)\footnotemark[4]}  & \mbox{$T_c$ (K)\footnotemark[5]}\\
\hline
\oLSMO 
& 380.8(2) & 3.5(2) &  370.3(3) &370(1) &  
1.75(2) &
0.309(2) & 0.310(6) 
& 371(2) & ~ \\

La$_{5/8}$(Ca$_{0.33}$Sr$_{0.67}$)$_{3/8}$MnO$_3$ & 
380.9(3) & 4.3(1) &  349.3(1)& 348.8(4) &  
1.73(2) &
0.299(2) & 0.294(3) & 349.7(4) 
& 350(1) \\

La$_{5/8}$(Ca$_{0.44}$Sr$_{0.56}$)$_{3/8}$MnO$_3$ & 
378.9(2) & 3.8(1) &  342.6(1)& 340.0(4)&  
1.80(1) &
0.311(5) &0.288(3) & 342.1(7) & 342.4(5) \\

La$_{5/8}$(Ca$_{0.5}$Sr$_{0.5}$)$_{3/8}$MnO$_3$ & 
380.7(1) & 3.3(1) &  335.5(3) &335(1)  &  
 1.85(2) &
0.309(3) & 0.305(6) & 333.4(7) & 333(2) \\

La$_{5/8}$(Ca$_{0.56}$Sr$_{0.44}$)$_{3/8}$MnO$_3$ & 
 380.3(2) & 3.6(1) &  335.0(3) & 331.9(4) &  
1.79(2) &
0.328(3) & 0.295(4) & 334.1(4) & 332.2(8) \\

La$_{5/8}$(Ca$_{0.67}$Sr$_{0.33}$)$_{3/8}$MnO$_3$ & 
379.6(2) & 3.3(1) &  316.8(2) & 313.6(4) &  
1.84(2) &
0.305(4) & 0.278(3) & 317.2(8) & 316.8(8) \\

La$_{5/8}$(Ca$_{0.78}$Sr$_{0.22}$)$_{3/8}$MnO$_3$ & 
379.5(2) & ~ &  ~ & 298.9(2) &
1.79(3)  &
 ~ &  0.256(5) &~ & 300.6(4) \\

La$_{5/8}$(Ca$_{0.89}$Sr$_{0.11}$)$_{3/8}$MnO$_3$ & 
379.7(2) & ~ & ~ & 286.9(2)&
1.85(2)&
 ~& 0.254(4)& ~ & 286.3(5) \\

\oLCMO & 
380.1(4) & 3.5(1) &  268.5(1) & 272(2) &  
1.84(5) &
0.211(4) & 0.231(8)  & 268.5(2) & 270.7(5) \\

La$_{0.75}$Sr$_{0.25}$MnO$_3$ & 
387.9(3) & ~ & ~ &  354(1) &
1.90(3) &
~ &  0.306(7) ~ & & 343(2) \\

La$_{0.8}$Sr$_{0.2}$MnO$_3$ & 
389.8(1) & ~ &  ~ &  335(2)&
2.00(2) &
  ~ &  0.315(7) & ~ & 310(3) \\

La$_{0.7}$Ca$_{0.3}$MnO$_3$ & 
385.1(3) & ~ & ~ & 272(1) &
1.82(3) &
 ~ & 0.218(6) & 248(1) & 251(1) \\
La$_{0.75}$Ca$_{0.25}$MnO$_3$ & 
388.1(2) & ~ & ~ & 252(3) & 
2.00(3) &
~ & 0.23(1) & 194.3(3) & 198(3) \\
\end{tabular}
\end{ruledtabular}
\footnotetext[1]{\hbox{Defined as  
$B_{\mu}^{\circ}\equiv   2\pi\kappa \,{^{55}\nu}_0 / \gamma_\mu$, 
where 
$\kappa$ is the empirical scaling constant 
between  $\nu_\mu(T)$ and ${^{55}\nu}(T)$ in the combined fit to Eq.\ \protect{\ref{eq:mlaw}}.}}
\footnotetext[2]{From combined $\mu$SR-NMR data.}
\footnotetext[3]{From NMR data only.}
\footnotetext[4]{Mean value, from recovery of the longitudinal muon polarization.}
\footnotetext[5]{Mean value, from the  rf response $(L(T)-L_0) / L_0$.}
\end{table*}

\begin{figure}
\includegraphics[width=\columnwidth]{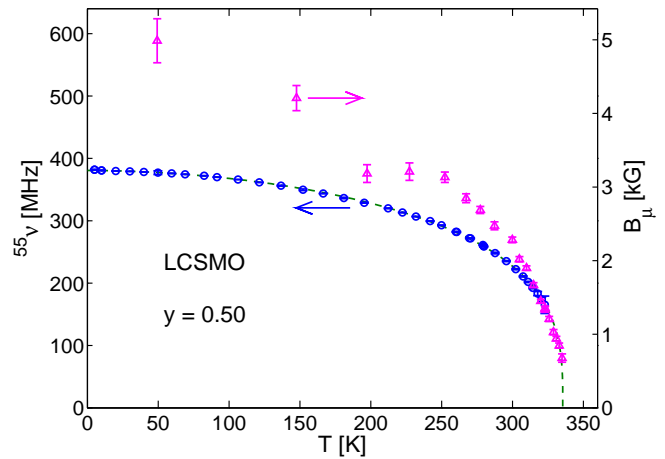}
\caption{\label{fig:combined_mlaw}(color online) $^{55}\nu$ (bullets) and 
$B_\mu$ (triangles) vs. $T$ in LCSMO $y=0.5$. The dashed line is a fit of the 
NMR and high-temperature $\mu$SR data to Eq.\ \ref{eq:mlaw}, as described 
in the text.}
\end{figure}

A typical fit of $^{55}\nu$ and $\nu_\mu$ vs.\ $T$ to the 
phenomenological 
function of Eq.\ \ref{eq:mlaw}, with a scaling constant 
$\kappa\equiv {^{55}\nu}(T) / \nu_\mu(T)$ 
as an additional free parameter, is illustrated in 
Fig. \ref{fig:combined_mlaw} for a 
representative compound ($y=0.5$). Here, the parameters $\alpha$ and 
$^{55}\nu_0$ are essentially determined by the 
$^{55}$Mn frequencies, while the values of $T_c^\star=T_c$ and the critical 
exponent $\beta$ mostly depend on the $\mu$SR data.
The best fit parameters of all the samples are summarized in 
Tab.~\ref{tab:mlaw_pars}. 
For comparison, the table 
also reports $T_c^\star=T_c$ and $\beta$ determined solely by the NMR data.
Despite the severe extrapolation of $^{55}\nu(T)$ for $T\to T_c$, the NMR 
estimate of the critical parameters reasonably agrees with the more 
precise $\mu$SR determination. 
This 
makes us confident about using
the extrapolation of the NMR frequency in the materials where muon precession 
data are not available (i.e.\ $y=0.78$, 0.89).

\begin{figure}
\includegraphics[width=\columnwidth]{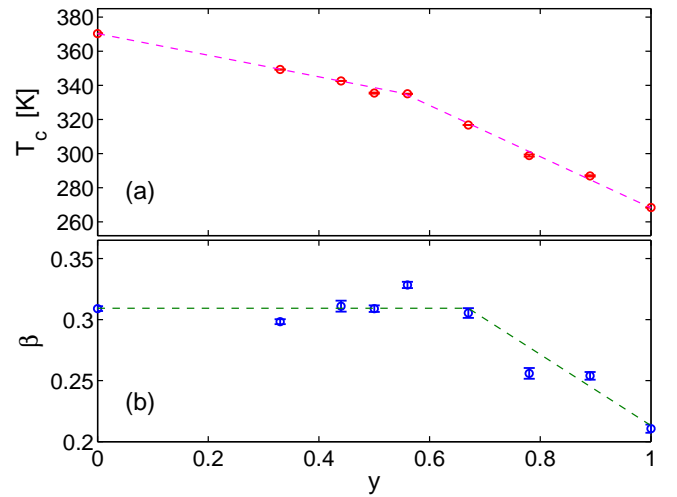}
\caption{\label{fig:beta_Tc_vs_y}(color online) 
$T_c=T_c^\star$ (a) 
and critical exponent $\beta$ (b) as a function of $y$ in 
the \oLCSMO\ series, determined from the combined fit of ${^{55}\nu}(T)$ and 
$\nu_\mu(T)$ ($y=0.78$, 0.89: fit of ${^{55}\nu}$ 
only) to  Eq.\ \ref{eq:mlaw}. The dashed lines are guides to the eye.
}
\end{figure}

The best fit values of $T_c^\star=T_c$ and $\beta$ across the LCSMO series 
are plotted vs.\ $y$ in Fig.~\ref{fig:beta_Tc_vs_y}. 
The Curie temperature shows a smooth dependence on $y$,  
slightly steeper for $y >  0.5$, but with no singularity at 
$y \approx 0.5$-0.6, where X ray diffraction detects a transition from 
the $R\bar 3c$ ($y < 0.5$) to the $Pnma$ space group.
{\redtwo
The exponent $\beta$ exhibits an approximately constant value
$\beta=0.31(1)$ with  increasing $y$ up to $y=2/3$, 
above which it decreases gradually,  down 
to $\beta=0.21(1)$ in the $y=1$ compound. 
The latter is 
in good agreement with the value reported in the literature for nearly 
optimally doped LCMO, determined by $\mu$SR. \cite{heffner_prb2001}
For reference, the critical exponent in 
LSMO was  determined as $\beta=0.37(4)$ by magnetometry. \cite{lsmo_MvsT} 
Our smaller absolute $\beta$ 
value at low $y$ is probably due to 
employing a fit to the empirical law of Eq.\ \ref{eq:mlaw} over the whole 
temperature range, instead of a proper analysis of the critical point.
However, the relative $\beta(y)$ dependence appears to be significant,  
also in view of best-fit $\alpha$ exponents nearly constant over the LCSMO
series (Tab.~\ref{tab:mlaw_pars}). 
The reduced value of $\beta$ found in LCMO reflects therefore} a steeper
temperature dependence of the order parameter close to the magnetic 
transition, which  however retains its continuous character. 
The relevance of $\beta$ for the nature of the magnetic transition is discussed in some detail in the next section. 

\section{Discussion and conclusions}

\label{sec:discussion}

In the isoelectronic 
\oLCSMO\ series at optimum hole doping, 
the magnetic transitions are second 
order at all compositions, as it is apparent from the vanishing muon 
precession frequencies at $T_c$. In the case of the end member \oLCMO, this 
agrees with a previous finding by $\mu$SR,  \cite{heffner2003} 
but contrasts with Adams {\it et al.},\cite{lynn} reporting a
finite limit value 
of the spin wave stiffness   
at  $T_c$ as in a first order transition.
In Ref.\ \onlinecite{lynn},  
however, the truncation effect is small, and the discrepancy 
in the transition order might be accounted for 
by a slight underdoping (see below), as indicated by a slightly lower 
$T_c$ in that sample. 
The Curie point 
decreases smoothly from 370~K ($y=0$) down to 268~K ($y=1$), 
with a  more marked $y$ dependence in the orthorhombic phases on the LCMO side, 
but without any apparent singularity 
in either $T_c$ or the critical exponent $\beta$
at the $R\bar 3c$-$Pnma$ space group transition (Fig.~\ref{fig:beta_Tc_vs_y}). 
The change in crystal symmetry, nominally occurring at $y\approx0.55$, is 
accompanied by a $y$ interval where the two phases coexist 
microscopically, and no crystal structure can be refined globally 
(Fig.~\ref{fig:2theta}).
The 
magnetic transition, however, is apparently  
insensitive to such structural inhomogeneities. This indicates that 
their spatial scale is much shorter than the  
coherence length of the conduction holes, so that the electronic motion 
averages the exchange coupling, which is controlled by the local lattice 
distortions,
into a single effective interaction.

In contrast, the transitions are 
truncated 
in all our underdoped samples, including LSMO. 
Notably, the first order transition of \uLSMO\ 
($T_c\approx 340$~K) takes place at a much higher temperature than 
the continuous transition in optimally doped LCMO (Fig.~\ref{fig:NMR_vs_T}). 
Our observation of 
a first order transition in high-T$_c$ LSMO, well beyond the insulator-metal 
phase boundary $x=0.17$,\cite{tokura_lsmo} 
challenges the currently accepted belief that the
transition order is essentially controlled by the size mismatch of the dopant 
cation. \cite{lynn, mossbauer, tricriticalLCSMO30}
Here, we provide evidence that the truncation phenomenon
is primarily an effect of underdoping and, possibly, also of overdoping. 

\begin{figure}
\includegraphics[width=\columnwidth]{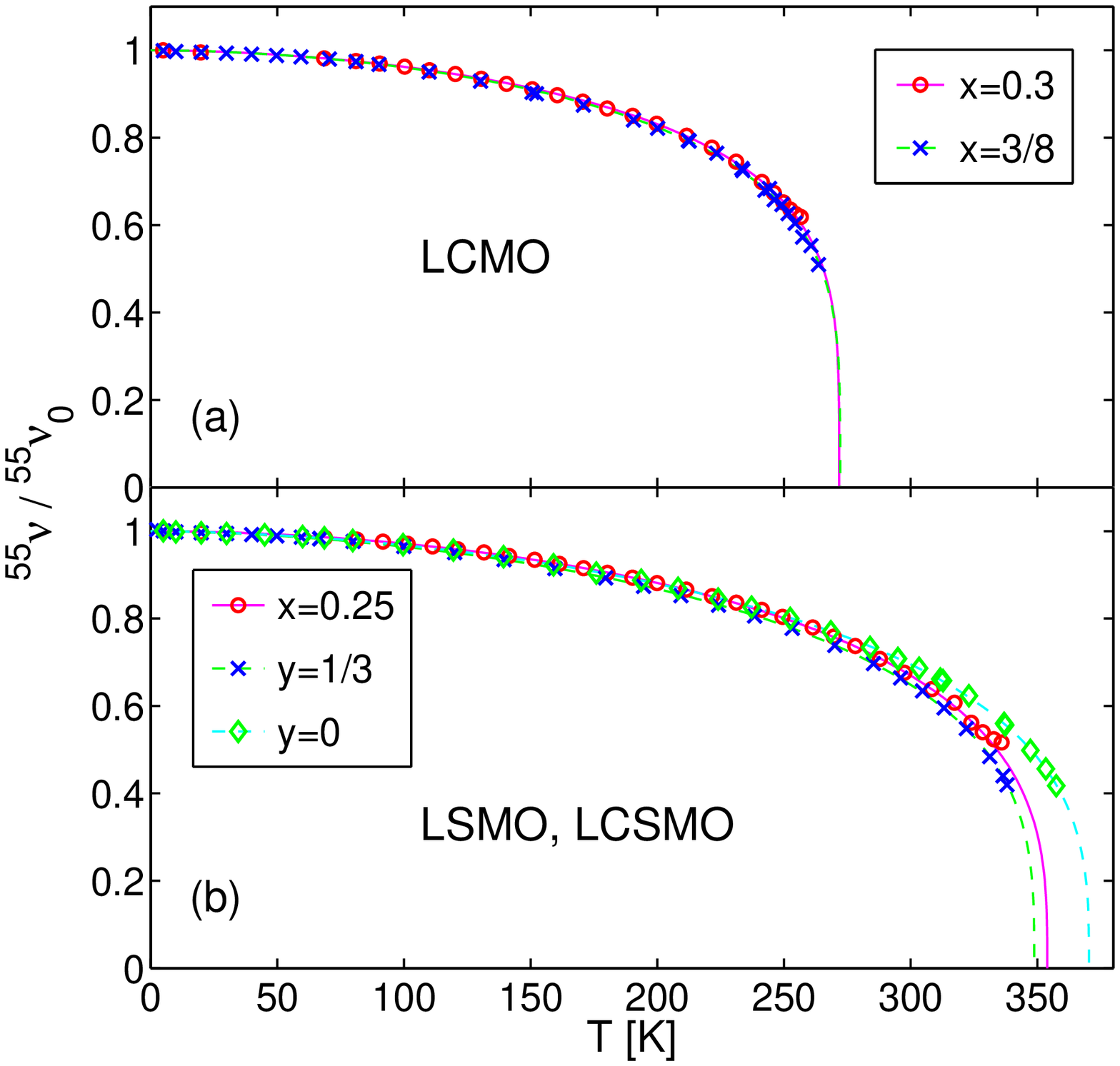}
\caption{\label{fig:nicescaling}(color online) Top: normalized $^{55}$Mn 
hyperfine field vs.\ $T$ in \uLCMO\ (bullets) and, for comparison,
 in \oLCMO\ (crosses).
Bottom:  the same 
quantity in \uLSMO\ (bullets),  compared with \oLSMO\ (diamonds) and 
La$_{5/8}$(Ca$_{0.33}$Sr$_{0.67}$)$_{3/8}$\-MnO$_3$ (crosses).
The lines overlaid to symbols are fits to 
Eq.\ \protect{\ref{eq:mlaw}} of combined NMR-$\mu$SR data 
(optimally doped compounds) or NMR data alone (underdoped LCMO and LSMO).
}
\end{figure}

Hole doping, on the other hand, seems to play a 
minor role in the 
temperature 
dependence of the order parameter, and hence in the effective exchange 
coupling.
The order parameter curves of \LCSMO\ exhibit in fact remarkable scaling 
properties vs.\ $x$, exemplified in  Fig.~\ref{fig:nicescaling}. 
The normalized $^{55}$Mn spontaneous frequencies $^{55}\nu(T)/^{55}\nu_0$ of 
optimally doped and underdoped ($x=0.3$) LCMO are compared in 
Fig.~\ref{fig:nicescaling}a. The overlap of the two curves is nearly perfect, 
if we except the truncation of  $^{55}\nu(T)$ at $T_c\approx 248$~K in the 
underdoped compound. The insensitivity of the order parameter curve and 
$T_c^\star$ on hole
doping $x$ in LCMO, in contrast to their strong dependence 
on the cation substitution 
$y$ in the optimally doped series (Fig.~\ref{fig:beta_Tc_vs_y}), suggests that 
the strength of the exchange interaction is primarily governed by 
distortions. Actually, the structural parameters in the $x=0.3$ sample 
(not shown) were found coincident within errors with those of LCMO $x=3/8$ 
(Fig.~\ref{fig:rietveld}), in accordance with their common behavior 
in $^{55}\nu(T)$.
The scenario emerging from the $x$ and $y$ dependence of the order parameter, 
respectively, in \LCMO\ and \oLCSMO, therefore
fits qualitatively into the original DE model, \cite{zener, anderson_hasegawa}
whereby the exchange coupling constant $J_{DE}$ 
is determined by the overlap of the 
Mn $3d$ with 
the ligand $2p$ wave functions via the Mn-O-Mn bond 
angle. 

The case of LSMO  is slightly more complex. The 
normalized $^{55}$Mn frequencies of the $x=0.25$ sample 
(Fig.~\ref{fig:nicescaling}b) significantly deviate from those of the 
optimally doped LSMO compound, and interpolate between the latter and 
the one of 
LCSMO ($y=0.33$), 
as also indicated by the extrapolated Curie temperatures $T_c^\star$ 
(Tab.~\ref{tab:mlaw_pars}) of these three materials. 
The Mn-O-Mn bond in the  strongly underdoped \uLSMO, however, 
exhibits a significantly larger distortion than in 
optimally doped LSMO (Fig.~\ref{fig:rietveld}).
This suggests that 
the  moderate variation of the order parameter curves
in these two 
samples is an indirect effect of doping, and is mostly 
due to the dependence of 
lattice distortions on the Sr concentration.

\begin{figure}
\includegraphics[width=\columnwidth]{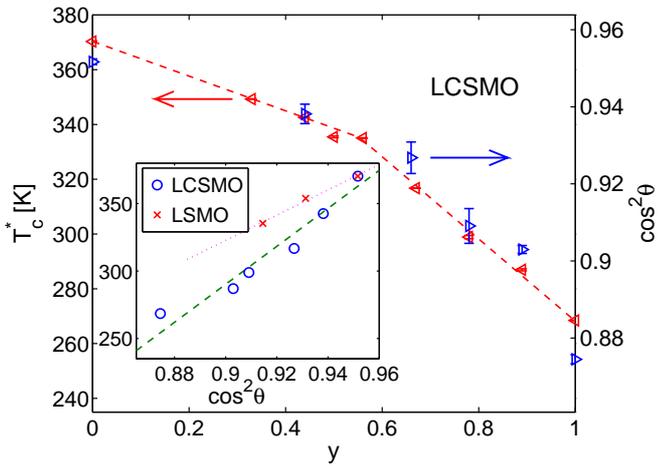}
\caption{\label{fig:thetaTcscale}(color online) 
Squared cosine of the mean Mn-O-Mn bond angle $\theta$ vs.\ $y$ in 
optimally doped LCSMO
(right triangles),   
overlaid with a linear scaling to 
$T_c=T_c^\star$ vs.\ $y$ (left triangles). The dashed line is the best-fit of 
$T_c^\star(y)$ to a broken line. Inset: $T_c^\star$ plotted vs.\   
$\cos^2\theta$ in optimally doped LCSMO (bullets) and in 
LSMO (crosses). 
}
\end{figure}

The existing correlation between $T_c^\star$  (hence, $J_{DE}$) and the mean 
Mn-O-Mn bond angle $\theta$ is illustrated in Fig.~\ref{fig:thetaTcscale},
plotting together, with a linear scale 
transformation, $T_c^\star$ and $\cos^2\theta$ data vs.\ $y$ in the \oLCSMO\
series.
The latter represents, to leading order, the bond angle dependence of
double exchange predicted by theoretical calculations,\cite{DE_MnOMn_theory}
and therefore it should be regarded as proportional to 
$J_{DE}$. 
Seemingly, the two quantities scale reasonably well with each other.
Their linear correlation 
in LCSMO is also apparent in the figure inset, showing $T_c^\star$ as a 
function
of $\cos^2\theta$. For reference, the inset also plots data points from the
underdoped LSMO samples, showing a qualitatively similar 
monotonic decrease
of $T_c^\star$ with decreasing $\cos^2\theta$. However, the latter 
deviate significantly from the LCSMO points with $y > 0$. 
The lack of a universal $T_c^\star(\theta)$ scaling law valid
for all compositions possibly indicates that other structural parameters 
are also at play in determining $J_{DE}$. For instance, the 
extra $A$-site cation disorder present in the LCSMO compounds might play a 
role in lowering their $T_c^\star$ as compared to LSMO, as already 
proposed by Rodriguez-Martinez {\it et al.} for other CMR 
manganites.\cite{attfield, attfield_98} 
Nevertheless, we believe that the view of an exchange coupling 
governed by distortions is substantially correct, 
although we have not succeeded to single out an effective distortion
parameter determining $J_{DE}$ univocally.

\begin{figure}
\includegraphics[width=\columnwidth]{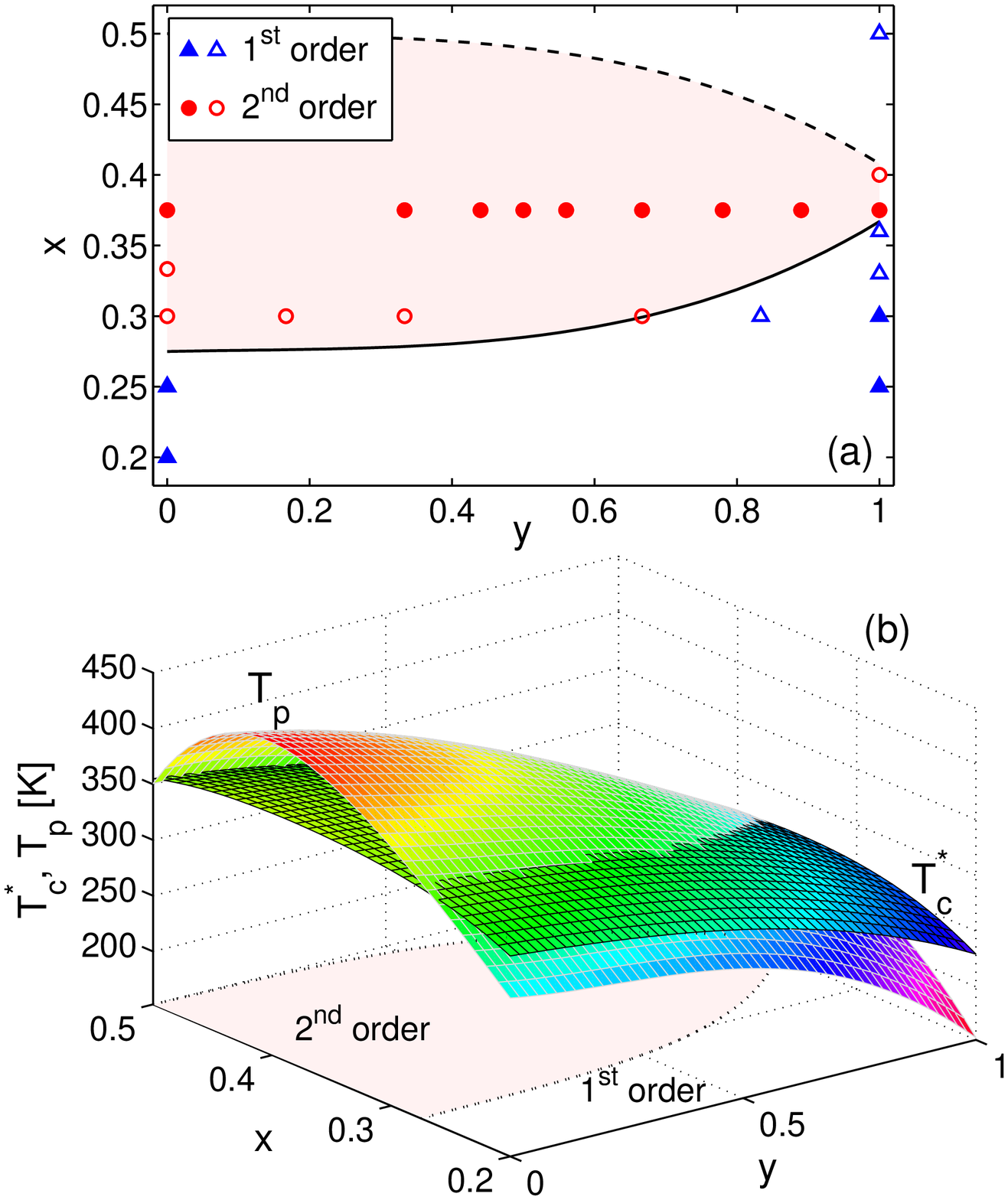}
\caption{\label{fig:phadia}(color online) Top: first- 
(triangles) and 
second 
order Curie transitions (bullets) in metallic \LCSMO\ as a function of 
$x$ and $y$.
Solid and open symbols 
mark samples studied in the present work and in the literature, 
respectively. The nominally \oLCMO\ sample  by Adams {\it et al.}
\protect{\cite{lynn}} 
is marked  assuming a slight underdoping ($x=0.36$) (see text).
Lines are a guess for the boundary between 
the two types of transitions. 
Bottom: guess for the $x$-$y$ dependence of the characteristic temperatures 
$T_p$ and $T_c^\star$ in LCSMO (see text). Second order transitions 
(shaded area in the $x$-$y$ plane) correspond to compositions 
whereby $T_p(x,y) > T_c^\star(x,y)$.
}
\end{figure}

The picture sketched 
above whereby the order of the 
magnetic transition is determined by charge doping, however, 
is clearly oversimplified.
A truncated $\nu(T)$ curve  indicates in fact the 
disruption of the DE metallic state 
 due to the formation of a polaron phase. \cite{lynn}
More realistically,
the relative stability of the two phases 
must also depend on the single-electron metallic bandwidth. The 
latter is closely related to the 
strength of the DE interaction, \cite{anderson_hasegawa} which in turn
is controlled by lattice distortions, according to the above arguments.
A tentative boundary 
in the charge-distortion space, separating first- from second-order Curie transitions, 
is plotted for the 
\LCSMO\ system in Fig.~\ref{fig:phadia}a, with $y$ as the implicit distortion 
parameter. 
In the figure we also include 
determinations of the order of the transition by other authors, 
obtained by $\mu$SR, \cite{heffner, heffner_prb2001}
$^{139}$La NMR, \cite{allodi_ca50}
 magnetometry, \cite{tricriticalLCSMO30, tricritical}
specific heat and thermal expansion measurements. \cite{tricritical, mira} 
The existence of a phase boundary across the whole $y$ range in the 
overdoping regime, where data at $y < 1$ are missing (dashed 
line), is just argued  on the basis of 
the strongly truncated transition observed in 
La$_{1/2}$Ca$_{1/2}$MnO$_3$.   \cite{allodi_ca50}
From the reported continuous transition in slightly overdoped LCMO ($x=0.4$), 
\cite{tricritical}
we also guess a certain asymmetry of the high- and low-doping (solid line) 
boundaries with respect to optimal doping $x=3/8$.

The prominent role of polarons in the truncated transitions of \LCSMO\ is 
indicated, among other evidence, by transport measurements, showing 
a semiconductive behavior of the electrical resistivity $\rho(T)$ above $T_c$ 
e.g.\ in underdoped LCMO, 
accompanied by a huge 
magnetoresistance effect in the vicinity of the transition. 
\cite{phadia_lcmo, rho_LCMO}
This contrasts with the transport properties of the LSMO compounds showing 
second order magnetic transitions, which exhibit shallower
magnetoresistance peaks at $T_c$ and resistivity maxima well above the Curie 
point. \cite{tomioka}
The metallic-like conductivity at the magnetic transition 
in 
nearly optimally doped LSMO 
indicates the 
absence of a polaron phase 
just above $T_c$. Indeed, the 
localization of polarons in \LSMO\ ($x=0.3$, $0.4$) has been detected by 
soft X ray
photoemission spectroscopy at temperatures $T_p > 450$~K, 
i.e.\ well above $T_c$. \cite{mannella} 
Combining these findings with data of ours and from the authors cited 
above, we are led to postulate the existence of a polaron localization 
temperature $T_p$ at all  metallic compositions in \LCSMO, with a 
$T_p(x,y)$ dependence similar to the one sketched in 
Fig.~\ref{fig:phadia}b.
The magnetic transition is then 
governed by critical spin fluctuations, as in \oLSMO, 
or driven by polarons, as 
in \uLCMO, depending on whether the
condition 
$T_p > T_c^\star$ or $T_p < T_c^\star$ is met, 
respectively.
The phase  boundary between first and second order transitions 
(Fig.~\ref{fig:phadia}a) corresponds therefore to 
the intersection of the $T_p(x,y)$ and $T_c^\star(x,y)$ surfaces. 

Second order transitions 
occur 
over hole-doping intervals that apparently 
narrow at increasing substitution of Sr with Ca. 
The end member \oLCMO, which is very close to the 
border with truncated transitions, is therefore  
virtually 
a singular point in the 
isoelectronic \oLCSMO\ series.
The near-singularity of optimally doped LCMO
is also signalled by a critical 
exponent  $\beta\approx 0.21$, from a fit to Eq.\ \ref{eq:mlaw}, 
in contrast to $\beta\approx 0.31$ in LSMO, 
indicating a steeper drop of the order parameter at $T_c$ in the 
former (Fig.~\ref{fig:beta_Tc_vs_y}b). 
Reduced $\beta\approx 0.25$, numerically 
coincident with the 
exponent of a second 
order transition at a tricritical point, \cite{huang_book}
were similarly 
 determined from magnetization measurements \cite{note_on_beta} 
in La$_{0.6}$Ca$_{0.4}$MnO$_3$  
by Kim {\it et al.\ }\cite{tricritical} and in underdoped
{La$_{0.7}$(Ca$_{2/3}$Sr$_{1/3}$)$_{0.3}$\-MnO$_3$} 
by Phan {\it et al.}\cite{tricriticalLCSMO30}
Those authors relate 
the tricritical point $\beta$ value to the crossing of 
the boundary between 
continuous and second order transitions in the $x$-$y$ phase diagram, and 
thus they implicitly identify the third phase at close equilibrium with 
the polaron phase 
responsible for the 
truncated  transitions.
We note however that $\beta$ does not exhibit any reduction 
in our underdoped LSMO samples showing a first order transition. In the latter,
 $\beta$ is intended as the fitting parameter to the phenomenological law of 
Eq.\ \ref{eq:mlaw}, denoting
the critical exponent of the virtual FM transition 
which would take place at $T_c^\star$, if the transition were not truncated. 
Although such a determination relies on an extrapolation, the fitted $\beta$
values seem reliable, especially in \uLSMO\ where the truncation effect is 
moderate ($T_c^{\star}-T_c\approx 11$~K). 
For reference, the fits  to Eq.\ \ref{eq:mlaw} of sole 
$^{55}$Mn NMR data from optimally doped LCSMO, based on similar extrapolations,
yield fitting parameters in good agreement with those determined by combined 
NMR and $\mu$SR data, covering temperatures up to $T_c$ 
(Tab.~\ref{tab:mlaw_pars}). 
Conversely, the fits of the truncated $^{55}\nu(T)$ curves in underdoped LCMO
yield the same reduced $\beta$ as in \oLCMO. 
In summary, we can confidently state that small $\beta$, close to the 
appropriate value for a tricritical point, are typical of the most distorted
manganites like LCMO,  
and have no
apparent direct relation  
with the vicinity  to a first order transition.
The nature of the tricritical transition point remains therefore unclear.

In conclusion, we have shown that the exchange coupling in metallic DE 
manganites, hence the one electron bandwidth, is primarily controlled by 
distortions and is nearly insensitive to hole doping. In contrast, 
the second vs.\ first
order of the Curie transition is determined by the interplay between 
bandwidth and 
band filling, where the latter apparently plays a dominant role.


\section*{ACKNOWLEDGEMENT}
The authors thank A.\ Amato (PSI) for assistance and helpful
discussion, and 
F.\ Licci and T.\ Besagni (IMEM-CNR, Parma) for help and 
precious suggestions in the sample synthesis.
The technical support of the Laboratory for Muon Spectroscopy 
and of the accelerator staff of the Paul Scherrer Institute are gratefully
acknowledged.


\appendix

\section{The hyperfine coupling of $^{55}$Mn}
\label{sec:coupling.nmr}

We comment here on the 
dependence of the $^{55}$Mn hyperfine field $B_{hf}$ on composition, 
and its relevance for 
the electronic state of the Mn ion in CMR manganites.

The hyperfine coupling of the $^{55}$Mn nuclei to the 
Mn electronic 
moments $\bm{S}$, of the form  
$\bm{B}_{hf} = g\mu_B{\cal  A}\VEV{\bm{S}}$  
(here ${\cal  A}$ is the 
mean
hyperfine coupling constant), 
is  known to be 
 negative and isotropic in metallic pseudocubic
manganites.  \cite{allodi_ca50, allodi_jmmm}
The isotropic character of ${\cal  A}$ is a manifestation of  
the so-called orbital liquid state of these systems, \cite{orbital_liquid}
which effectively leads to averaging out all the 
anisotropic (i.e.\ pseudodipolar) hyperfine terms typical of non 
closed-shell ions as \mnt, due to the fast
motion of the delocalized $e_g$ holes.
As a result of motional averaging, comparatively sharp single peaks are 
detected in metallic manganites, in place of the complex and very broad 
$^{55}$Mn spectra typical of underdoped insulating compositions. \cite{kapusta}
The negative sign and the magnitude of ${\cal  A}$ demonstrate the 
dominant origin of the hyperfine coupling of $^{55}$Mn from the Fermi contact 
interaction with the core $s$ wave functions, which are spin-polarized by the 
outer $3d$ shell. 
This core polarization mechanism contributes an isotropic and essentially 
single-ion hyperfine coupling 
term, yielding a spontaneous field proportional to 
the on-site total
spin. Its coupling coefficient 
${\cal  A}_{cp}$ approximately equals a value of -10~T/$\mu_B$ 
in iron-group transition metals, roughly independent of the ionic species 
within ten percent accuracy.
\cite{RadoSuhl}. 

The zero-temperature spontaneous resonance frequency $^{55}\nu_0$ 
is found to depend only 
on the hole concentration $x$ according to Eq.\ \ref{eq:straight}, 
irrespective of the dopant species.
This finding qualitatively agrees with 
an essentially on-site hyperfine coupling of $^{55}$Mn, 
whereby $^{55}\nu_0(x)$ is proportional to the local spin-only Mn 
electronic moment $gS\mu_B$, which is controlled in turn by the hole 
concentration, $g\VEV{S(x)}= 4-x$. 
Such a $S(x)$ dependence 
is well established 
experimentally e.g.\ by neutron diffraction. \cite{xiong}
Nevertheless, 
the proportionality of 
$^{55}\nu_0(x)$ to $S(x)$ would imply
slope $a$ and  intercept $b$ (Eq.\ \ref{eq:straight}) in a ratio $a/b=-1/4$,
which is not verified experimentally.
The $b=403(2)$~MHz parameter would correspond to 
a hyperfine coupling constant ${\cal A}=b/(4\mu_B {^{59}\gamma}\,/\,2\pi) 
\approx 
-9.6~\tesla/\mu_B$ in the hypothetical 
$x=0$ compound, in loose agreement with the known core polarization value. 
The $a=-60(2)$~MHz parameter, 
though of the correct 
negative sign, 
is however smaller than $b/4$ by nearly a factor of two. 
The reduced magnitude of $a$ therefore indicates the presence of 
extra $x$-dependent isotropic and positive contributions to the $^{55}$Mn 
hyperfine field, which partly cancel the on-site core polarization term.

A well-known positive term, present in both insulating and metallic magnetic 
oxides, is given by the 
super-transferred hyperfine coupling via oxygen. \cite{owen_taylor_1973} 
Its leading contribution consists in the polarization of the local core 
$s$ wave functions by a $d$ orbital of a neighboring Mn ion, mediated by
the ligand wave functions. \cite{owen_taylor_PR} 
Oxygen-assisted charge transfer 
into outer 4$s$ Mn wave functions also contributes with the same 
sign. \cite{orbach}
In either case, the $d$ orbitals participating in the hyperfine field transfer
are the singly occupied $e_g$ wave functions overlapping to the ligand $p$ 
orbitals. \cite{owen_taylor_PR, orbach}    

We can estimate the super-transferred hyperfine coupling in CMR manganites
from its experimental determination in the isostructural
La(Ni$_{1-x}$Mg$_x$)$_{0.5}$Mn$_{0.5}$O$_3$ perovskite. 
There, each $Ni^{2+}$ ion ($t_{2g}^6e_g^2$ configuration) 
contributes 0.63~T to the 
hyperfine field at the nucleus of a neighboring Mn$^{4+}$ ion, yielding an 
overall 
transferred field of 3.8~T at $^{55}$Mn in LaNi$_{0.5}$Mn$_{0.5}$O$_3$. 
\cite{sonobe_asai}  
From calculations, similar values were estimated for the  transferred 
hyperfine interaction from Mn$^{2+}$ ($t_{2g}^3e_g^2$) to Mn$^{2+}$, 
for instance 
in MnO. \cite{owen_taylor_PR} 
For the supertransferred hyperfine field at $^{55}$Mn in CMR 
manganites, we assume the same value as in LaNi$_{0.5}$Mn$_{0.5}$O$_3$, scaled 
by the populations of the $e_g$ orbitals in the two classes of compounds.  
In mixed valence LCMO and LSMO, the orbital liquid state implies an occupation 
probability of $(1-x)/2$ for the two Mn $e_g$ wave functions.
We predict therefore in the 
present compounds a transferred hyperfine field $B_{st}=(1-x)B_{st}^{(0)}$,
with $B_{st}^{(0)}\approx 1.9$~T . Its actual 
value might be smaller owing to the non-collinearity of the Mn-O-Mn bond, 
reducing the overlap between the Mn and the ligand wave functions, 
in contrast to the La(Ni$_{1-x}$Mg$_x$)$_{0.5}$Mn$_{0.5}$O$_3$ case.

Such an estimate for $B_{st}^{(0)}$ is to be compared with the observed 
$^{55}\nu_0(x)$ dependence. Following the above arguments, the total
 hyperfine field at $^{55}$Mn is written as
\begin{equation} 
B_{hf}=\mu_B {\cal A}_{cp}(4-x)+ {\cal B}(1-x)
\label{eq:Bhf}
\end{equation} 
where the second term on 
the right hand side 
has the form of a transferred hyperfine field.
From Eq.\ \ref{eq:straight} and \ref{eq:Bhf}, with 
$^{55}\nu_0 = - ^{55}\gamma B_{hf}$, we obtain
${\cal A}_{cp}= -\nicefrac{1}{3}\,(a+b)\,/\, ({^{55}\gamma}\mu_B )=
-10.9(1)$~T/$\mu_B$ 
and
${\cal B} = \nicefrac{1}{3}\,(4a+b)\,/\, {^{55}\gamma}=5.7(2)$~T. 
The latter is larger than
the estimated $B_{st}^{(0)}$ by approximately a factor of 3. 

The discrepancy between ${\cal B}$ and $B_{st}^{(0)}$ may be reconciled by 
assuming a further positive $x$-dependent hyperfine contribution. 
To our knowledge, the only isotropic and positive hyperfine term possible in a 
metal, in addition 
to supertransfer, is the direct contact interaction 
with conduction electrons in outer $s$ shells, 
like e.g.\ in metallic copper. 
In CMR manganites, a $4s$-wave component with metallic character can only 
arise from a slight hybridization of the valence $3d$:$e_g$ wave functions. 
A similar $3d$-$4s$ wave admixture was actually 
proposed in superconducting  cuprates 
by Mila and Rice, \cite{milarice} in order to explain the anomalous 
Knight shifts and spin-lattice relaxation rates of Cu nuclei. 
Like the former contribution, this $3d$-$4s$ mixing term must be proportional 
to the $3d$:$e_g$ electronic population, and therefore 
it must be written as 
$B_{4s}=(1-x)B_{4s}^{(0)}$, formally identical to $B_{st}$.
Eq.\ \ref{eq:Bhf} is then reformulated into
\begin{equation} 
B_{hf}=\mu_B {\cal A}_{cp}(4-x)+ (B_{st}^{(0)}+B_{4s}^{(0)})(1-x)
\label{eq:Bhf_st4s}
\end{equation} 
where the $3d$-$4s$ term $B_{4s}^{(0)}= {\cal B}- B_{st}^{(0)}$ is estimated 
on the order of 4~T.


\section{The muon coupling}
\label{sec:coupling.muon}

We comment here on the anomalous coupling of muons to the local Mn moments in 
metallic manganites, as it appears from 
the temperature dependence of the spontaneous frequencies 
$\nu_\mu$ and $^{55}\nu$  and from the 
longitudinal relaxations of the two probes of magnetism.
The precession frequency $\nu(T)$ and longitudinal 
relaxation rates $\lambda(T)\equiv T_1^{-1}(T)$
of a local probe of magnetism
are simply proportional to the thermodynamic expectation value of the 
electronic spins and to the generalized spin susceptibility at the Larmor 
frequency $\chi''(\omega_L)$, respectively, under the implicit 
hypothesis of  temperature independent   
hyperfine and dipolar couplings.
In most cases, however, such a temperature independence is just postulated. 
The deviation of $\nu_\mu(T)$ from $^{55}\nu(T)$ 
(Fig. \ref{fig:combined_mlaw})
actually demonstrates that in metallic manganites the 
coupling of either probe varies with temperature.

The coupling of $^{55}$Mn is given to leading order by
the Fermi contact interaction with the inner $s$ shells via the core 
polarization mechanism, \cite{RadoSuhl}
which is both isotropic and virtually independent on the ion surroundings. 
Transferred contributions from the neighboring spins 
are isotropic as well and smaller by an order of magnitude (see  
Appendix \ref{sec:coupling.nmr}). Although  in 
principle 
transferred hyperfine terms may be sensitive to the local 
structure, their residual variance is expected to be a small fraction, 
at most in the order of 1\% of the total hyperfine field $B_{hf}$.
Experimentally, 
$B_{hf}(0)$ is found to be
independent of the dopant ion, in spite of the different degree of 
distortion in the LSMO, LSCO, and LCSMO families (Fig.~\ref{fig:NMR_nu0}).
Muons, on the other hand, are coupled to the electronic moments
by {\redtwo 
the dipolar interaction plus, possibly, a usually smaller hyperfine 
contact term. The former is anisotropic and strongly 
dependent on the muon stopping site, which 
is determined in turn} by the electrostatic potential minima within the 
crystal cell. Thus, 
the local field at the muon is sensitive to both spin reorientations and 
structural changes. 

The above arguments demonstrate that coupling variations in the order of 
several tens percent, as those observed experimentally, may be ascribed 
solely to $\mu$SR.
Since a spin reorientation below $T_c$ has been ruled out by 
neutron diffractions data, \cite{lynn}
the anomaly of $\nu_\mu(T)$ must then be related to the multiplicity 
and relative stability of the muon sites.
The muon precession patterns well below $T_c$ actually indicate the presence 
of 
two stopping sites nearly degenerate in the internal field, plus 
extra sites whose overdamped precessions are not detected and give rise to 
missing transverse fractions. The recovery of 
simple damped cosine waveforms 
at higher temperature, with
the full transverse amplitude and relatively reduced inhomogeneous linewidths,
then points to the onset of a thermally activated hopping of the muon 
across its sites. If hopping takes place at a much faster rate than the 
precession 
frequencies, it effectively averages the various precessions into a single 
motionally narrowed line. 
In the presence of missing transverse fractions at low temperature and, 
possibly, 
temperature dependent branching ratios among the different sites, 
the transition from a hindered to 
a free motion of the muon may as well produce the step 
in its average dipolar coupling witnessed by the anomaly in $\nu_\mu(T)$
(inset of Fig.~\ref{fig:mu_freq_sigma}). 

The mobility of the implanted muon makes questionable its effectiveness 
as a probe of the dynamics of the electronic spins. In a magnetically 
ordered phase, 
muon hopping drives in fact a random time-dependent modulation of the large 
static dipolar field exerted by the electronic moments  at the muon. 
The spectral component at $\omega_L$ of such a field modulation 
contributes to the longitudinal muon spin relaxation. 
The relative importance  
of such a mechanism vs.\ the relaxation channel due to proper spin 
fluctuations is clarified  by the comparison with $^{55}$Mn spin lattice 
relaxations. 

\begin{figure}
\includegraphics[width=\columnwidth]{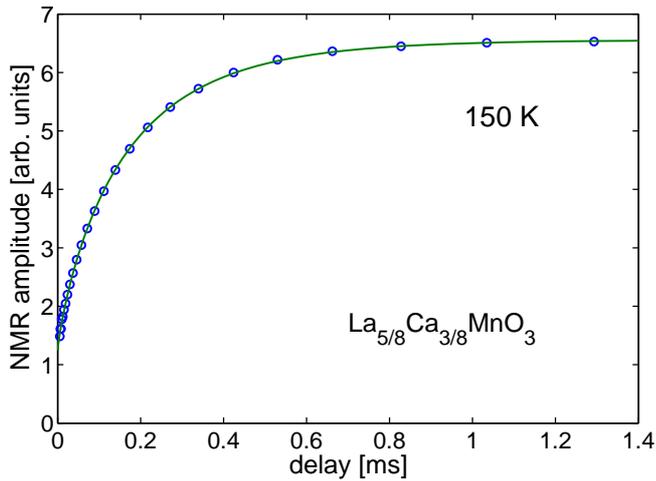}
\caption{\label{fig:t1_nmr}(color online)
 Recovery from saturation of the $^{55}$Mn NMR signal amplitude in 
optimally doped LCMO at 150~K, as a function of the
delay between the saturating pulse train and the spin echo detection.
The solid line is the best fit to a stretched exponential function 
(see text).}
\end{figure}

The recovery from saturation of the $^{55}$Mn nuclear magnetization $M_z$ is 
plotted vs.\ time in Fig.~\ref{fig:t1_nmr} for a typical compound 
(\oLCMO) at an intermediate temperature (150~K, approximately 0.6~$T_c$). 
The recovery curve is fitted to a stretched 
exponential, $M_z(t)=M_0(1-\exp(-(t/T_1)^\beta) + M_1$, with stretching 
exponent 
$\beta=0.8$ and mean relaxation rate $T_1^{-1}\approx 6\,10^3$~s$^{-1}$ 
(here $M_1$ is a residual unsaturated signal component). 
In the present case of NMR signal from domain walls, however, the 
dominant relaxation channel is due to an extrinsic mechanism, namely the 
mesoscopic fluctuations of the 
domain walls themselves. \cite{weger}
Their coupling to the nuclei is proportional to the domain wall enhancement 
$\eta_w$, which is strongly inhomogeneous across a wall, whence
 the non-exponential form of the recovery curves. 
These magnetic fluctuations are suppressed by an applied field saturating the 
magnetization, which typically decreases $T_1^{-1}$ by orders of 
magnitude. \cite{T1inLSMO}
Therefore, the above experimental value of $T_1^{-1}$ is a large overestimate
of its intrinsic value, arising from the spin excitations in the bulk of 
the domains. 
Anyway, we scale the measured relaxation rates of the two probes for their 
couplings to the electronic spins, in a ratio $^{55}\nu(0)/\nu_\mu(0) \ge 4$.
From the experimental spin lattice relaxation of $^{55}$Mn, we predict an
unmeasurably small muon 
relaxation rate $\lambda=T_1^{-1}\times (\nu_\mu(0)/^{55}\nu(0))^2 
\approx 4~10^2$~s$^{-1}$.  \cite{note_raman_process}
Such a value is to be compared to the smaller muon rate $\lambda_s$ in this 
compound, equal to 6~$10^4$~s$^{-1}$ at this 
temperature (Fig.~\ref{fig:mu_T1}b).
We conclude
therefore that the muon relaxation rates $\lambda_s$ and (a fortiori) 
$\lambda_f$ are both determined by the fluctuations of the local field at the 
muon induced by its own motion, 
rather than the fluctuations of the electronic spins.

The dominance of an extrinsic relaxation mechanism for the muon spin 
  perfectly explains
the lack of a critical relaxation peak at $T_c$ (Fig.~\ref{fig:mu_T1}b), 
if one excepts the broad bump in $\lambda_s$.
The comparison between NMR and $\mu$SR also proves that 
several conclusions drawn in the  
literature on $\mu$SR in CMR  manganites should be 
reconsidered. 
For instance, 
the bimodal decay of the longitudinal muon polarization was regarded by 
Heffner {\it et al.}\ as evidence for {\redtwo
two time scales in the 
spin dynamics of LCMO.\cite{heffner2000}
Nevertheless, it is now clear that the spatial 
inhomogeneity revealed by the longitudinal muon relaxation 
is rather related to a corresponding inhomogeneity in the 
motion dynamics of the muon. A thermally activated muon hopping in the 
presence of a broad distribution of potential barriers may be the 
 clue to both the non-exponential decays and their temperature evolution.
The outline of such a 
dynamical model, however, is outside the scope of this article. 
}



\begin{thebibliography}{}
\bibitem{note_on_beta}
The discrepancy between $\beta\approx 0.25$ 
in e.g.\ Ref.\ \protect{\onlinecite{tricritical}} and $\beta\approx 0.21$ in 
our \oLCMO\ sample is consistent with a similar difference between the 
determinations of
$\beta$ in nearly optimally doped LSMO, respectively, by 
magnetometry and NMR-$\mu$SR, where our techniques systematically detect
smaller values.
\end{thebibliography}
\end{document}